\documentclass[twocolumn,epj]{webofc}
\usepackage[varg]{txfonts}   
\usepackage{xcolor}
\usepackage{hyperref}
\hypersetup{
 linkbordercolor=red,
 citebordercolor=red,
 urlbordercolor=red
}
\graphicspath{{./pics/}}
\newif\ifcomment
\newif\ifarx
\arxtrue
\newcommand{\co}[1]        {\relax}
\newcommand{\pp}           {\mbox{pp}}
\newcommand{\Aa}           {\mbox{A--A}}
\newcommand{\PbPb}         {\mbox{Pb--Pb}}

\newcommand{\pPb}          {\mbox{p--Pb}}

\newcommand{\pt}           {\ensuremath{p_{\mathrm{T}}}}
\newcommand{\ptm}          {\ensuremath{p_{\mathrm{T,1}}}}
\newcommand{\pts}          {\ensuremath{p_{\mathrm{T,2}}}}
\newcommand{\snn}          {\ensuremath{\sqrt{s_{\mathrm{NN}}}}}

\newcommand{\avNcoll}      {\ensuremath{\left<N_\mathrm{coll}\right>}}
\newcommand{\etam}         {\ensuremath{\eta_\mathrm{1}}}
\newcommand{\etas}         {\ensuremath{\eta_\mathrm{2}}}
\newcommand{\etal}         {\ensuremath{\eta_\mathrm{lab}}}
\newcommand{\etac}         {\ensuremath{\eta_\mathrm{cms}}}
\newcommand{\Dphi}         {\ensuremath{\Delta\varphi}}
\newcommand{\Deta}         {\ensuremath{\Delta\eta}}
\newcommand{\RpPb}         {\ensuremath{R_\mathrm{pPb}}}
\newcommand{\Rfb}          {\ensuremath{R_\mathrm{FB}}}
\newcommand{\yl}           {\ensuremath{y_\mathrm{lab}}}
\newcommand{\yc}           {\ensuremath{y_\mathrm{cms}}}
\newcommand{\ptt}          {\ensuremath{p_{\mathrm{T, trig}}}}
\newcommand{\pta}          {\ensuremath{p_{\mathrm{T, assoc}}}}
\newcommand{\Ntrko}        {\ensuremath{N^{\mathrm{offline}}_{\mathrm{trk}}}}
\newcommand{\dNdeta}       {\ensuremath{\mathrm{d}N_\mathrm{ch}/\mathrm{d}\eta}}
\newcommand{\dNdetal}      {\ensuremath{\mathrm{d}N_\mathrm{ch}/\mathrm{d}\etal}}
\newcommand{\Nch}          {\ensuremath{N_{\rm ch}}}
\newcommand{\Ntracks}      {\ensuremath{N_{\rm tracks}}}
\newcommand{\kzero}        {\ensuremath{{\mathrm K}^{0}_{\mathrm S}}}
\newcommand{\pbar}         {$\rm\overline{p}$}
\newcommand{\allpi}        {$\pi^{\pm}$}
\newcommand{\allk}         {K$^{\pm}$}
\newcommand{\allp}         {p(\pbar)}
\newcommand{\lmb}          {\ensuremath{\Lambda}}
\newcommand{\almb}         {\ensuremath{\bar{\Lambda}}}
\newcommand{\alll}         {\lmb(\almb)}
\newcommand{\ptref}        {\ensuremath{p^{\mathrm{ref}}_{\mathrm{T}}}}
\newcommand{\sigpp}        {\ensuremath{\sigma_{\mathrm{pp}}}}
\newcommand{\sigppb}       {\ensuremath{\sigma_{\mathrm{pPb}}}}

\newcommand{\Fig}[1]       {Fig.~\ref{#1}}
\woctitle{LHCP 2013}
\begin{document}
\title{\vspace{-0.2cm}First results from \pPb\ collisions at the LHC}
%
%

\author{Constantin Loizides\inst{1}\fnsep\thanks{\email{cloizides@lbl.gov}}}

\institute{Lawrence Berkeley National Laboratory, 1 Cyclotron Rd, Berkeley, CA 97420, USA}

\abstract{The first results from \pPb\ collisions at \snn = 5.02 TeV are discussed.}
\maketitle
\section{Introduction}
\label{intro}
Proton--lead (\pPb) collisions are an integral part of the nuclear program at the 
Large Hadron Collider~(LHC). Their study provides the reference for the \PbPb\ data
to disentangle initial from final state effects, as well as the potential to address 
the partonic structure of matter at low parton fractional momenta (small-$x$)~\cite{Salgado:2011wc}.

The experimental results reported in these proceedings are obtained
in a short low-luminosity run performed in September 2012~(with about $1/{\rm\mu}$b recorded by each experiment), 
and a longer high-luminosity run in January/February 2013~(with about $30/$nb recorded by ATLAS and CMS,
about $10/$nb by ALICE and about $2/$nb by LHCb).
The setup of the beams, which is constrained by the two-in-one magnet design of the LHC imposing the same magnetic rigidity 
of the beams, consisted of protons at 4 TeV energy\co{ circulating in the negative $z$-direction} and of $^{208}_{82}$Pb ions 
at $82\times4$ TeV energy\co{ circulating in the positive $z$-direction}. 
This configuration produced collisions at $\snn=5.02$~TeV in the nucleon--nucleon centre-of-mass system, 
shifted in rapidity relative to the laboratory system by $\Delta y_{\rm NN}=0.465$ in the direction of the proton beam.
For clarity, the rapidity~($y$) as well as the pseudorapidity~($\eta$) are sometimes denoted as 
as $\yl$ and $\yc$, as well as $\etal$ and $\etac$.

To investigate the role of nuclear effects in \pPb\ collisions it is desirable to study experimental observables as a
function of centrality of the collision. In nucleus--nucleus~(\Aa) collisions this is typically achieved by relating 
intervals of measured multiplicity~(or energy) distributions~(that correspond to certain fractions of the inelastic 
cross-section) to an average number of nucleon--nucleon collisions~(\avNcoll) via a Glauber model~\cite{Abelev:2013qoq}.
In \pPb\ collisions, however, the correlation between multiplicity and collision geometry is less strong than in \Aa\ 
collisions, and more importantly dynamical biases introduced by the multiplicity estimation can strongly affect the 
observables under study~\cite{Morsch:2013xra}. Therefore, so far\co{ until centrality in \pPb\ is better understood}, the experimental results are either 
reported for minimum-bias collisions (where $\avNcoll=208\,\sigpp/\sigppb=7.0\pm0.6$, with interpolated $\sigpp=70\pm5$mb~\cite{ALICE:2012xs} 
and measured $\sigppb=2.07\pm0.07$b~\cite{LHCb-CONF-2012-034,CMS:2013rta}), 
or as a function of multiplicity, i.e.\ in selected intervals of 
a measured multiplicity or energy distribution without relating to centrality explicitly.
In the latter case, the selected intervals are typically characterized by the corresponding average charged-particle 
multiplicity at midrapidity. Potential biases introduced by the event selection can be studied by varying the underlying 
multiplicity or energy distribution.

The results presented at the conference include the measurements of the charged-particle pseudorapidity~\cite{ALICE:2012xs} 
and transverse momentum~($\pt$) distributions~\cite{ALICE:2012mj}, results on dijet~\cite{Chatrchyan:2014hqa} and
J/$\psi$~\cite{Abelev:2013yxa,Aaij:2013zxa} production, multiple results on long-range correlations of charged particles using 
two-particle~\cite{CMS:2012qk,Abelev:2012ola,Aad:2012gla} and four-particle~\cite{Aad:2013fja,Chatrchyan:2013nka} correlation 
analysis techniques, as well as results on identified particle $\pt$ distributions~\cite{Chatrchyan:2013eya,Abelev:2013haa}.
\ifarx
These, complemented by recent measurements on identified two-particle correlations~\cite{Abelev:2013wsa}, as well as
the measurement of the average $\pt$ of charged particles as a function of multiplicity at midrapidity~\cite{Abelev:2013bla}, 
will be discussed in the following.
\fi

\section{Unidentified charged particles}
\label{spectra}
\begin{figure}[tbh!f]
\centering
\includegraphics[width=6.5cm,clip]{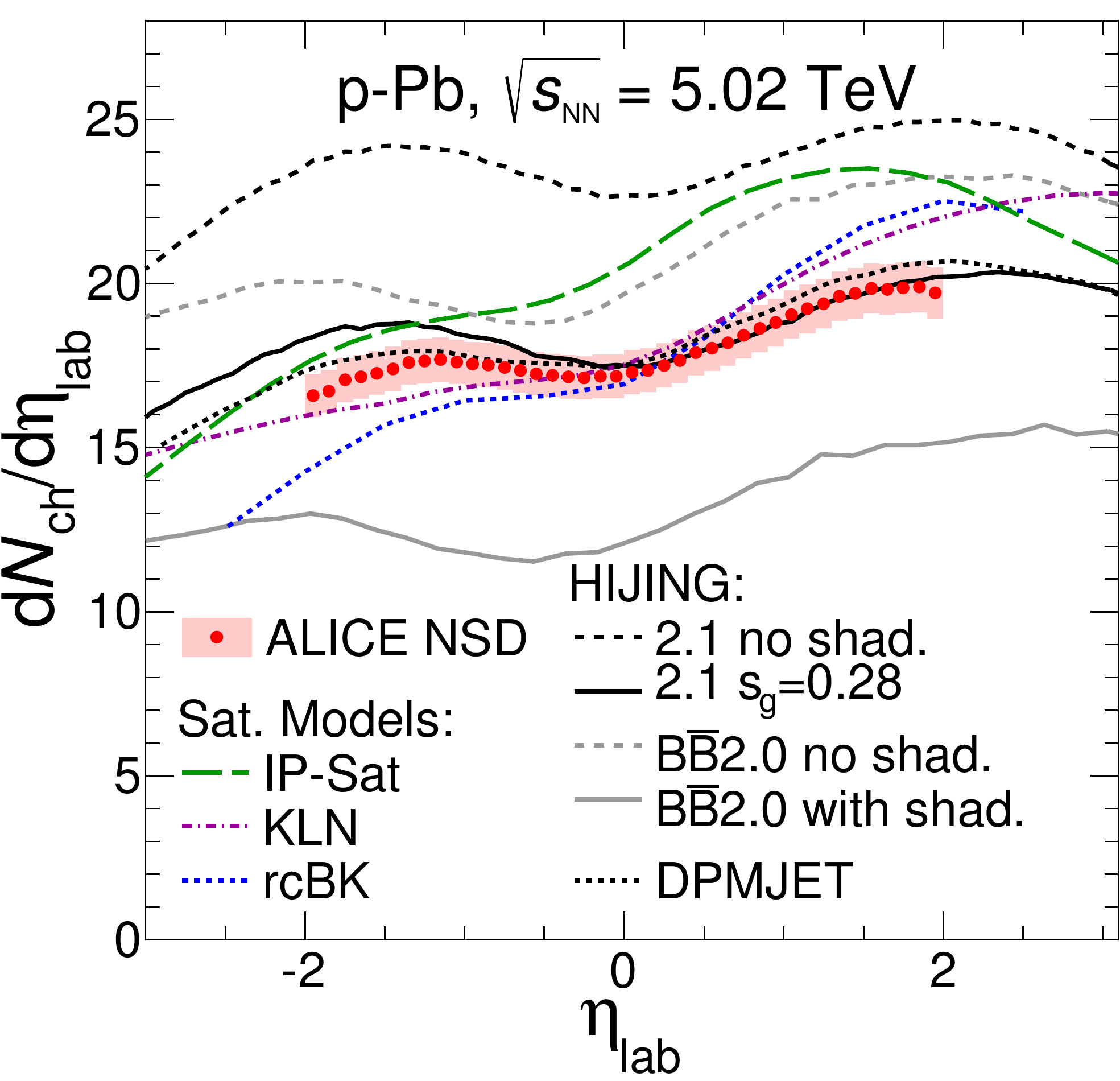}
\caption{
Pseudorapidity density of charged particles measured in the laboratory system~($\eta_{\rm lab}$) for non-single 
diffractive \pPb\ collisions at $\snn=5.02$~TeV compared to model predictions~\cite{ALICE:2012xs}.
}
\label{fig:dNdeta}
\end{figure}

The measurement of the charged-particle density provides constraints to improve the understanding of particle production 
and the role of initial state effects in QCD at small-$x$~\cite{ALICE:2012xs}.
The data are normalized to non-single diffractive collisions and reported in the laboratory system~(\Fig{fig:dNdeta}).
\begin{figure}[tbh!f]
\centering
\includegraphics[width=7cm,clip]{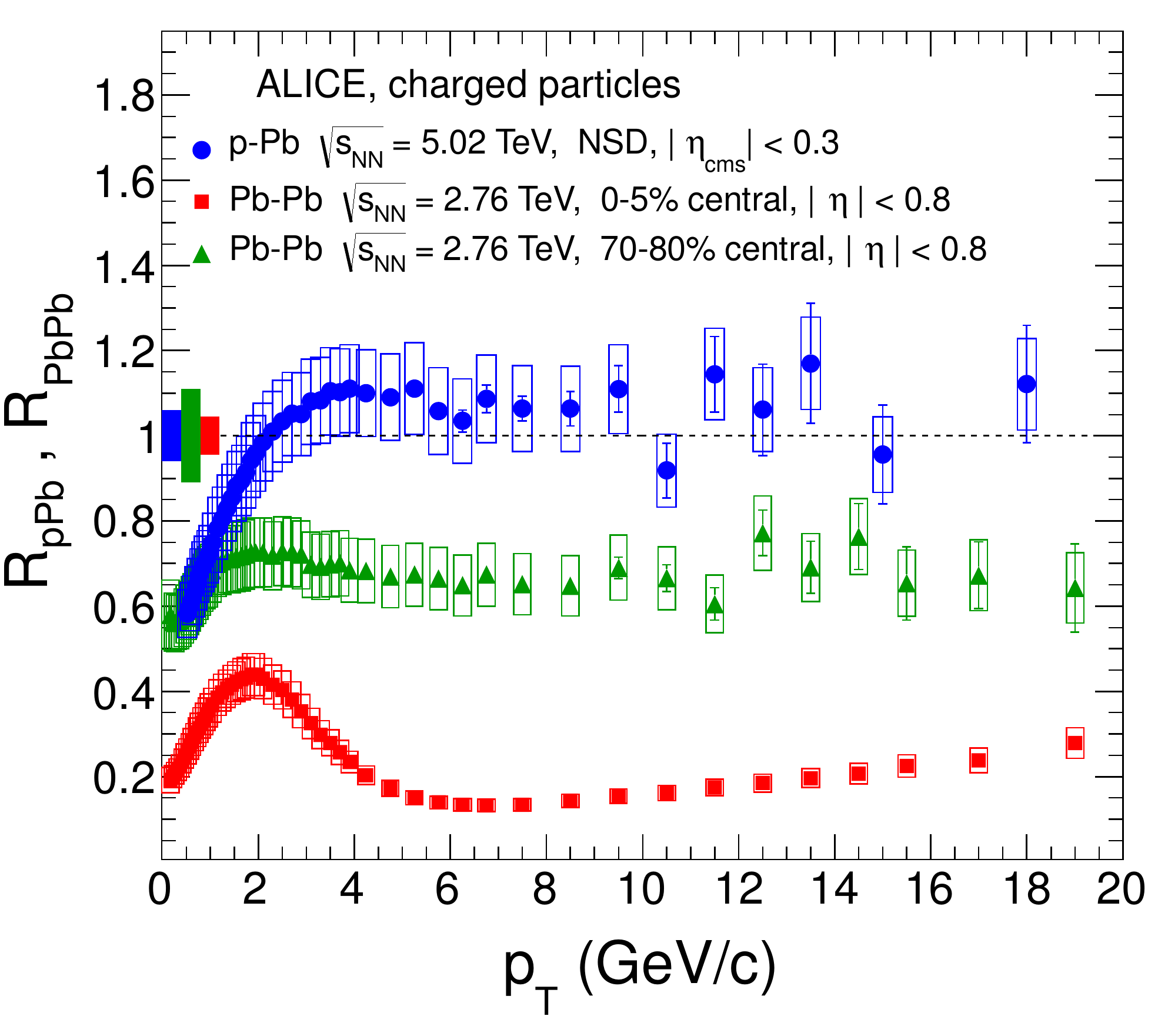}
\caption{
Nuclear modification factor for charged particles as a function of $\pt$ in $|\eta_{\rm cms}|<0.3$ in non-single 
diffractive \pPb\ collisions at $\snn=5.02$~TeV compared to data in $|\eta|<0.8$ in central~(0--5\%)
and peripheral~(70--80\%) \PbPb\ collisions at $\snn=2.76$~TeV~\cite{ALICE:2012mj}.}
\label{fig:dNdpt}
\end{figure}
The systematic uncertainty of the measurement is about 3.8\%. 
It is dominated by the uncertainty on the normalization, which is obtained by requiring that not all of the 
nucleon--nucleon collisions~(as for example modeled in the \mbox{DPMJET}~\cite{Roesler:2000he} generator) are single-diffractive.
The charged-particle pseudorapidity density at midrapidity in the laboratory system~($|\etal|<0.5$) is found to be
$17.4\pm0.7$, while the corresponding density in the nucleon--nucleon centre-of-mass system~($|\etac|<0.5$) to be $16.8\pm0.7$.
The measured distribution is compared to various model predictions~(references can be found in \cite{ALICE:2012xs}) that broadly 
can be characterized as either two-component or saturation models. The two-component models combine perturbative QCD processes 
with soft interactions, and may include nuclear modification of the initial parton distributions. 
The saturation models typically employ coherence effects to reduce the number of soft gluons available 
for particle production below a given energy scale.
The comparison with the data shows that most of the calculations predict the measured distribution to within 20\%, 
with a tendency that the the saturation models exhibit a steeper $\etal$ dependence than the data~(see also~\cite{Albacete:2013ei}).

\begin{figure}[tbh!f]
\centering
\includegraphics[width=8.0cm,clip]{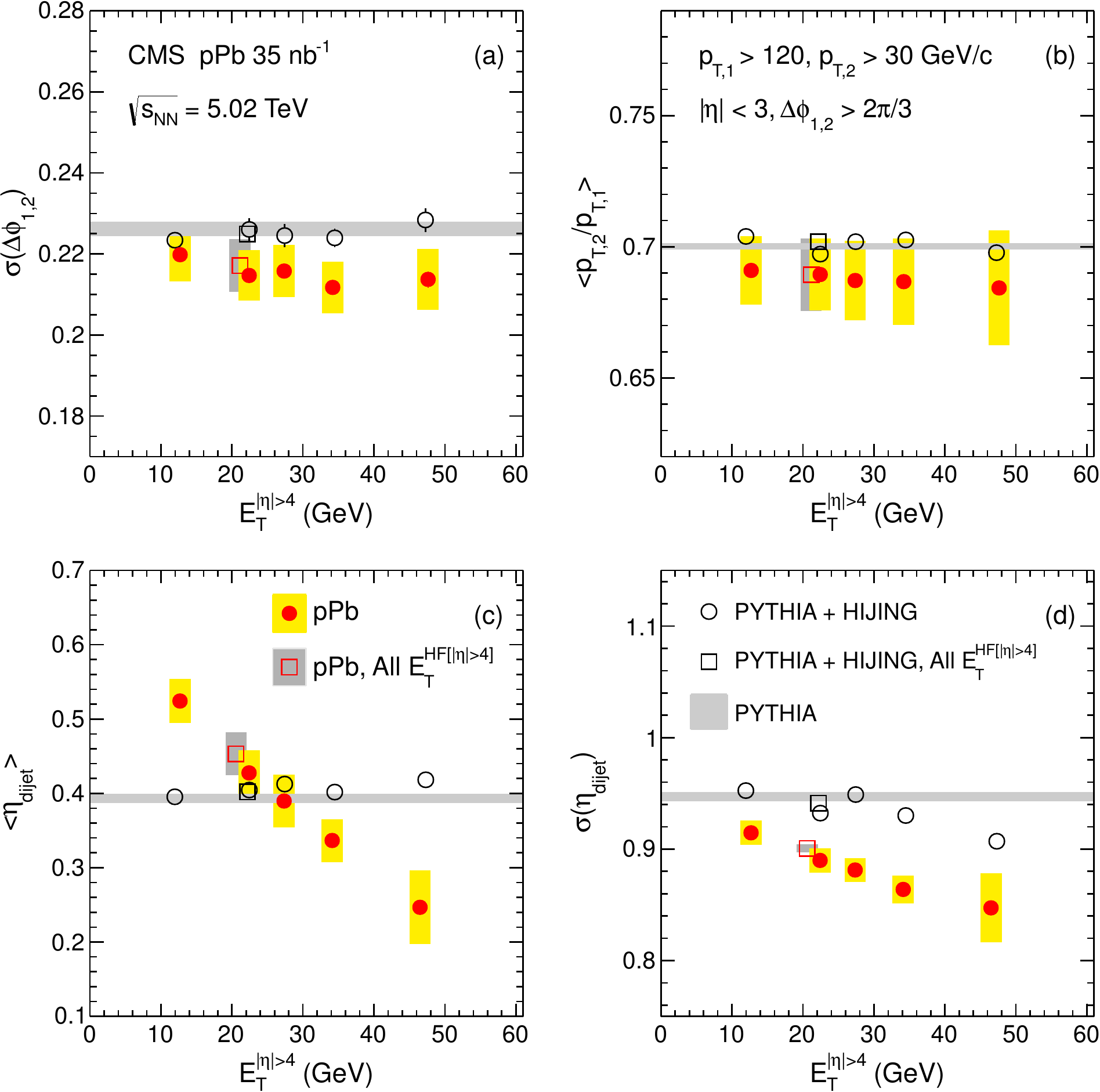}
\caption{
Dijet properties measured from jet pairs selected by requiring $\ptm>120$~GeV/$c$ for the leading 
and $\pts>30$~GeV/$c$ for the subleading jet, as well as an azimuthal angular difference of $\Delta\varphi_{\rm 1,2}>2\pi/3$ 
for jets reconstructed with the anti-$k_{\rm T}$ algorithm with $R=0.3$ in $\etal<3$~\cite{Chatrchyan:2014hqa}. 
The results are presented as a function of forward energy in \pPb\ collisions, 
compared to PYTHIA+HIJING \pPb\, as well as minimum bias PYTHIA \pp\ 
simulations~(shown as a blue band) at $\snn=5.02$~TeV.
(Top panels)~Width of the azimuthal distribution (left) and average transverse momentum ratio~(right) 
between subleading and leading jet.
(Bottom panels)~Average~(left) and width~(right) of the dijet pair pseudorapidity in the laboratory system.}
\label{fig:dijets}
\end{figure}

Further information on particle production are provided by the charged particle $\pt$ distributions, which are 
measured in $0.5<\pt<20$~GeV/$c$ for 3 ranges of $\etac$ near midrapidity normalized to non-single diffractive~(NSD) 
collisions~\cite{ALICE:2012mj}. The systematic uncertainty of the measurement is about 8--10\% including the uncertainty 
on the normalization. The spectra seem to soften with increasing pseudorapidity, although the effect
is of the similar magnitude as the systematic uncertainty.
It is found that most models that describe the $\etal$ distribution, like the DPMJET or HIJING models, 
have difficulties in describing the $\pt$ distributions. One exception is a calculation
from EPOS v3~\cite{Werner:2013tya}, which includes parton saturation and a hydrodynamical evolution.
Nuclear effects are usually quantified by the ratio of the yield extracted in \pPb\ collisions relative to that in \pp\ 
scaled by $\avNcoll$, which is expected to be unity in absence of nuclear effects. Since there are no \pp\ data at 
$\sqrt{s}=5.02$~TeV, the \pp\ reference is constructed by interpolating \pp\ data at 2.76 and 7 TeV~\cite{Abelev:2013ala}.
Using this reference, the nuclear modification factor, $\RpPb$, at $|\etac|<0.3$ is found to be consistent with 
unity for $\pt$ above 2 GeV/$c$, showing that there are no strong nuclear effects present in NSD~($\avNcoll\approx7$) \pPb\ 
collisions~(\Fig{fig:dNdpt}). Consequently, the measurement demonstrates that the high-$\pt$ suppression observed in central~(0--5\%, 
$\avNcoll\approx1700$) and peripheral~(70--80\%, $\avNcoll\approx16$) \PbPb\ collisions at $\snn=2.76$~TeV~\cite{Abelev:2012hxa} 
is not due initial-state, but rather due to final-state interactions.
It is interesting to note that the suppression in \PbPb\ is present already in 70--80\% collisions, where $\avNcoll$ is only 
twice as large as in NSD \pPb\ collisions. Therefore, final state effects could indeed play a role in more 
central \pPb\ collisions~(see Sections~\ref{correlations} and \ref{pidspectra}).

\section{Dijet production}
\label{dijets}
Complementary information is provided by a measurement of dijet production using 
an integrated luminosity of $18.5$/nb~\cite{Chatrchyan:2014hqa}. High-energy jets are reconstructed
with the anti-$k_{\rm T}$ algorithm~\cite{Cacciari:2008gp} for a resolution parameter of $R=0.3$ in $\etal<3$, 
using combined information from tracking and calorimetry.
Dijet pairs are selected by requiring $\ptm>120$~GeV/$c$ for the leading and $\pts>30$~GeV/$c$ for the subleading jet.
Then, the azimuthal angle correlations between the two jets~($\Delta\varphi_{\rm 1,2}$),
the dijet momentum balance~($\ptm/\pts$), and the mean and width of the dijet pseudorapidity distributions~($\frac{\etam+\etas}{2}$)
are measured for $\Delta\varphi_{\rm 1,2}>2\pi/3$ as a function of forward calorimeter transverse energy~(approximately
spanning a range of $\avNcoll$ between 5 to 15).
The data are compared to PYTHIA simulations representing \pp\ collisions, and to \pPb\ simulations using HIJING, 
where dijet pp events from PYTHIA are embedded~(\Fig{fig:dijets}).
The width of the azimuthal angle difference distribution and the dijet momentum ratio is not sensitive to the forward
activity of the collision, and comparable to the same quantity obtained from the simulations, confirming that the observed 
dijet asymmetry in \PbPb\ collisions~\cite{Aad:2010bu,Chatrchyan:2011sx} is not originating from initial state effects.
The pseudorapidity distribution of the dijet system, however, changes strongly with increasing forward calorimeter activity
in the nucleus direction, which indicates that the hard scattering process and the underlying event are strongly correlated.
The effects are much stronger than expected from the depletion of low-$x$ partons in nuclear Parton Distribution 
Functions~(nPDFs)~\cite{Eskola:2009uj}, which are found to describe the minbias \pPb\ collisions reasonably well\cite{Eskola:2013aya}.

\section{J/$\psi$ production} 
\label{jspis}
J/$\psi$ production in proton--nucleus collisions is expected to be sensitive to several initial and final state effects related 
to the presence of cold nuclear matter, such as the suppression of low-$x$ gluons and initial state energy loss~\cite{Albacete:2013ei}.
Results in \pPb\ collisions are available for inclusive J/$\psi$ in \mbox{$2<\yc<3.5$} and \mbox{$-4.5<\yc<-3.0$} 
using about 10.8/nb~\cite{Abelev:2013yxa} and for the first time separately for prompt J/$\psi$ and J/$\psi$ from b hadron decays in
\mbox{$1.5<\yc<4.0$} and \mbox{$-5.05<\yc<-2.5$} using about 1.6/nb~\cite{Aaij:2013zxa}. 
To obtain the nuclear modification factor, the \pp\ reference at forward rapidity is constructed by interpolating available lower and higher beam 
energy data with various functional forms~\cite{ALICE:2013spa}. 

\begin{figure}[tbh!f]
\centering
\ifarx
\includegraphics[width=7.1cm,clip]{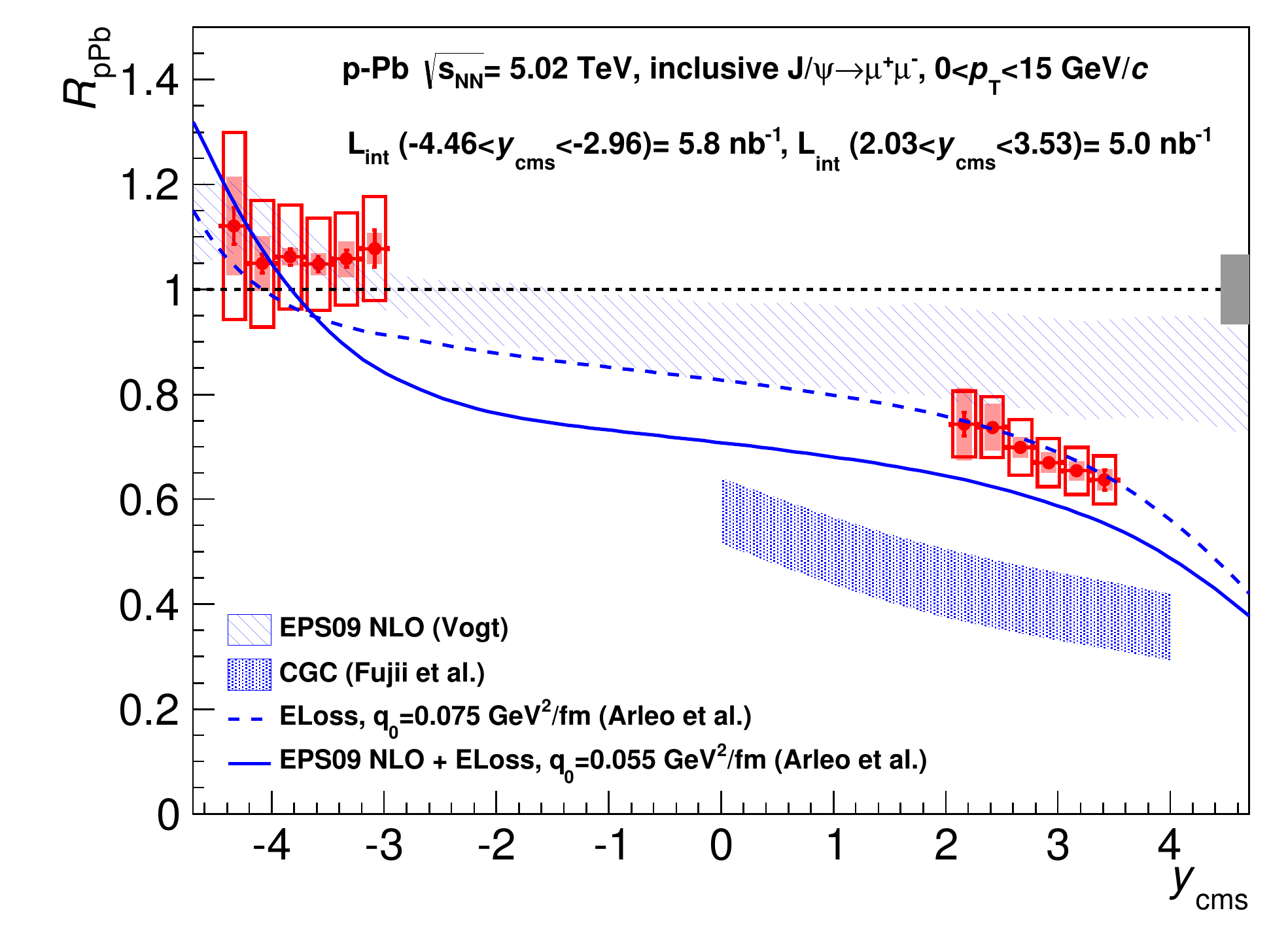}
\else
\includegraphics[width=6.8cm,clip]{RpA_RAp_Y_121213}
\fi
\caption{
Nuclear modification factor for inclusive J/$\psi$ as a function of $\yc$ in $0<\pt<15$~GeV/$c$ in NSD
\pPb\ collisions at $\snn=5.02$~TeV compared to calculations~\cite{Abelev:2013yxa}.}
\label{fig:jspirpa}
\end{figure}

At forward rapidity the inclusive J/$\psi$ production is suppressed~(with a mild rapidity dependence), 
compared to the backward rapidity~(\Fig{fig:jspirpa}).
\begin{figure}[tbh!f]
\centering
\ifarx
\includegraphics[width=7.3cm,clip]{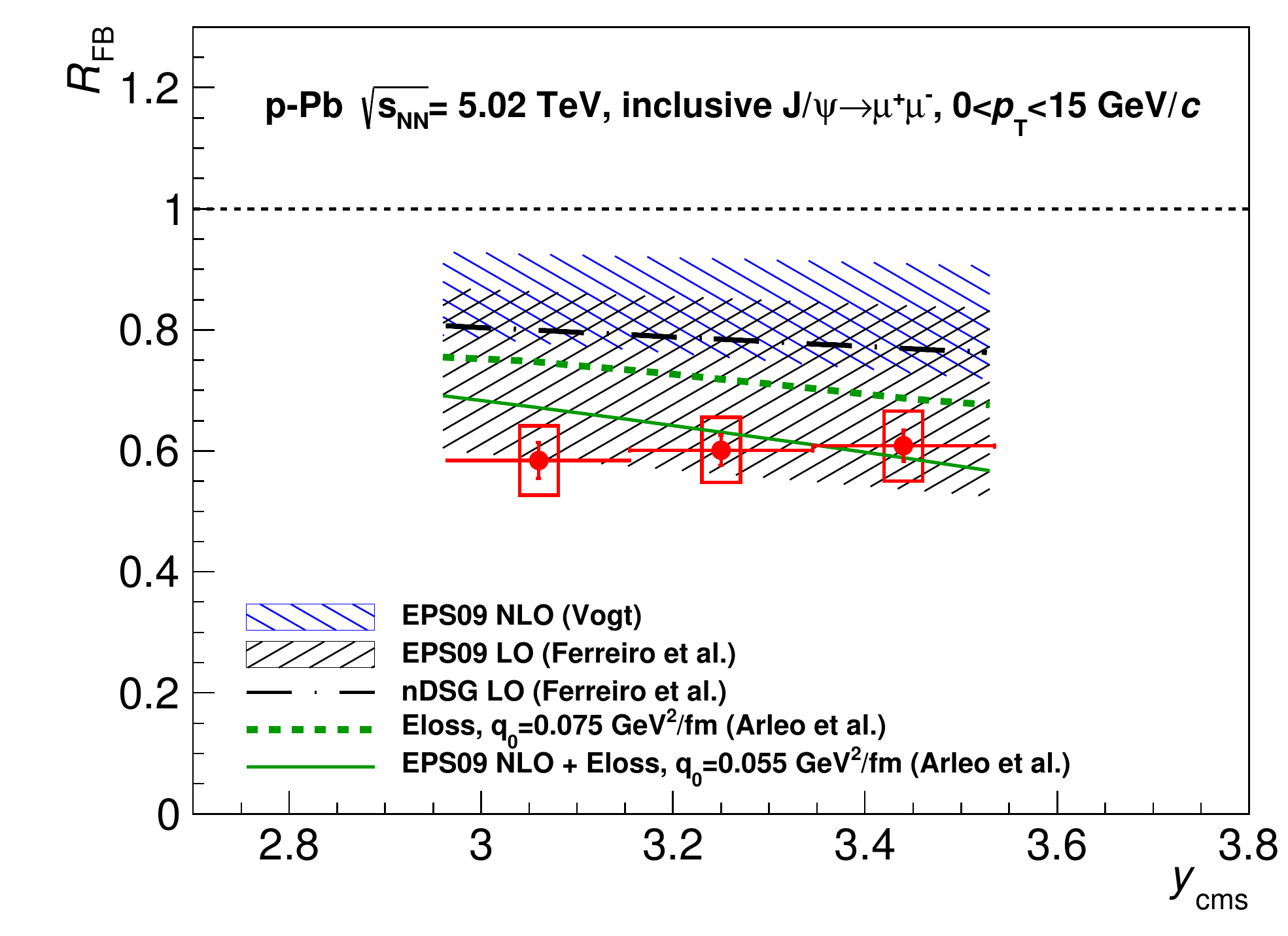}
\else
\includegraphics[width=7.0cm,clip]{RFB_Y_210813}
\fi
\caption{Forward-to-backward ratio for inclusive J/$\psi$ as a function of $\yc$ in $0<\pt<15$~GeV/$c$ 
in NSD \pPb\ collisions at $\snn=5.02$~TeV compared to calculations~\cite{Abelev:2013yxa}.}
\label{fig:jpsifby}
\end{figure}
\begin{figure}[tbh!f]
\centering
\ifarx
\includegraphics[width=7.3cm,clip]{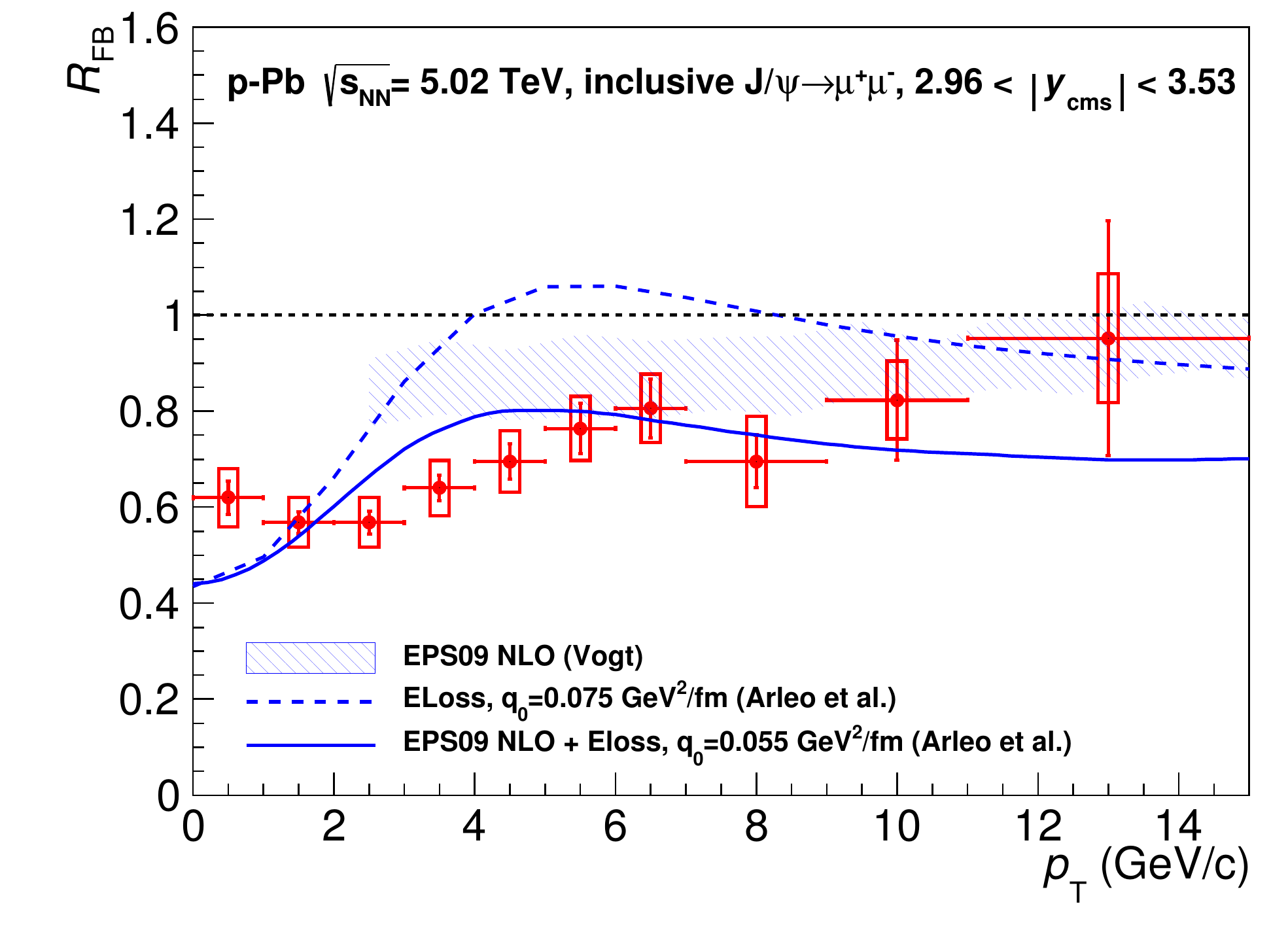}
\else
\includegraphics[width=7.0cm,clip]{RFB_pt_210813}
\fi
\caption{Forward-to-backward ratio for inclusive J/$\psi$ as a function of $\pt$ in $3.0 < \yc < 3.5$ in 
NSD \pPb\ collisions at $\snn=5.02$~TeV compared to calculations~\cite{Abelev:2013yxa}.}
\label{fig:jpsifbpt}
\end{figure}
\begin{figure*}[htb!f]
\centering
\includegraphics[width=5.2cm,clip]{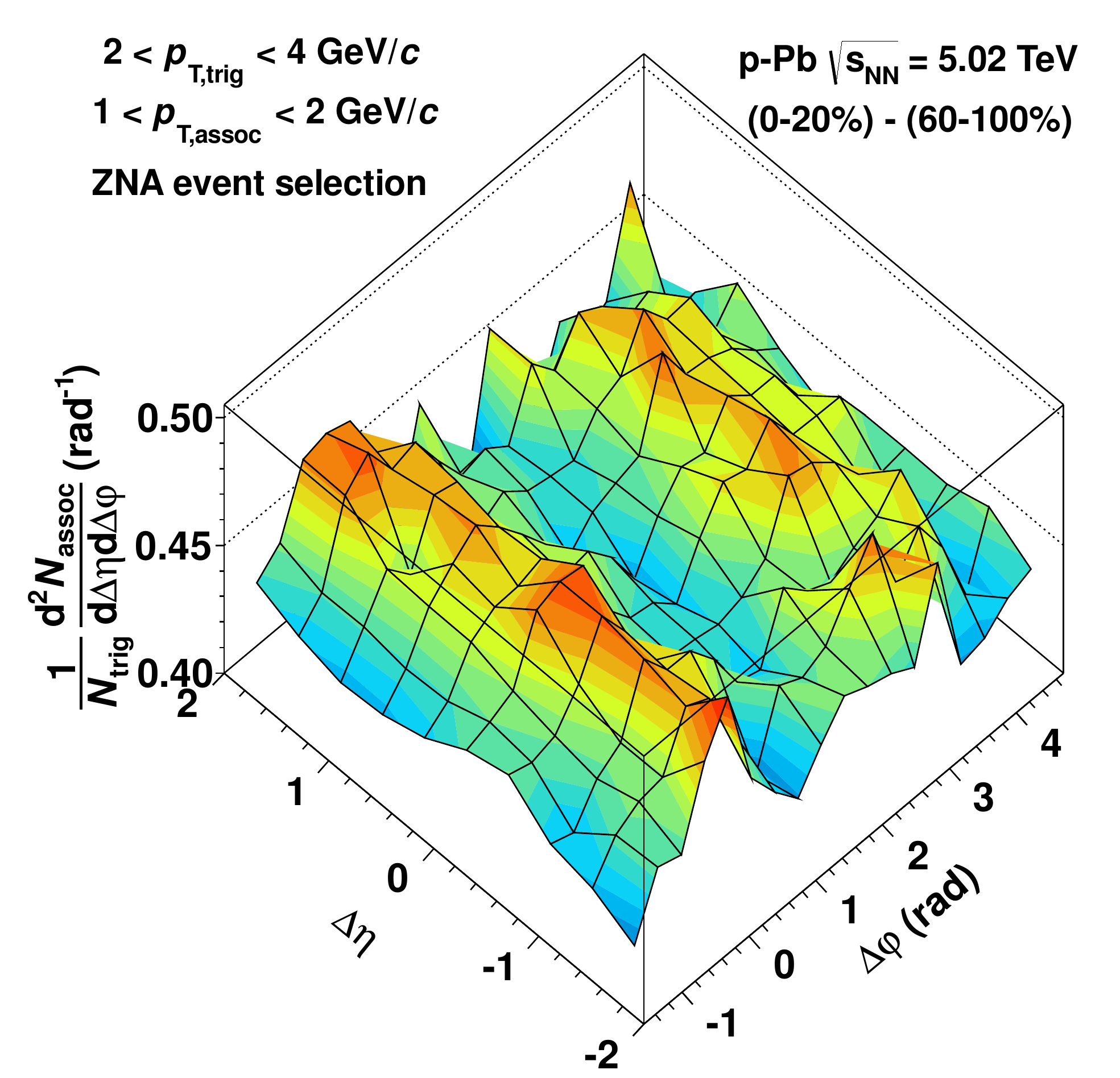}
\hspace{0.4cm}
\includegraphics[width=5.2cm,clip]{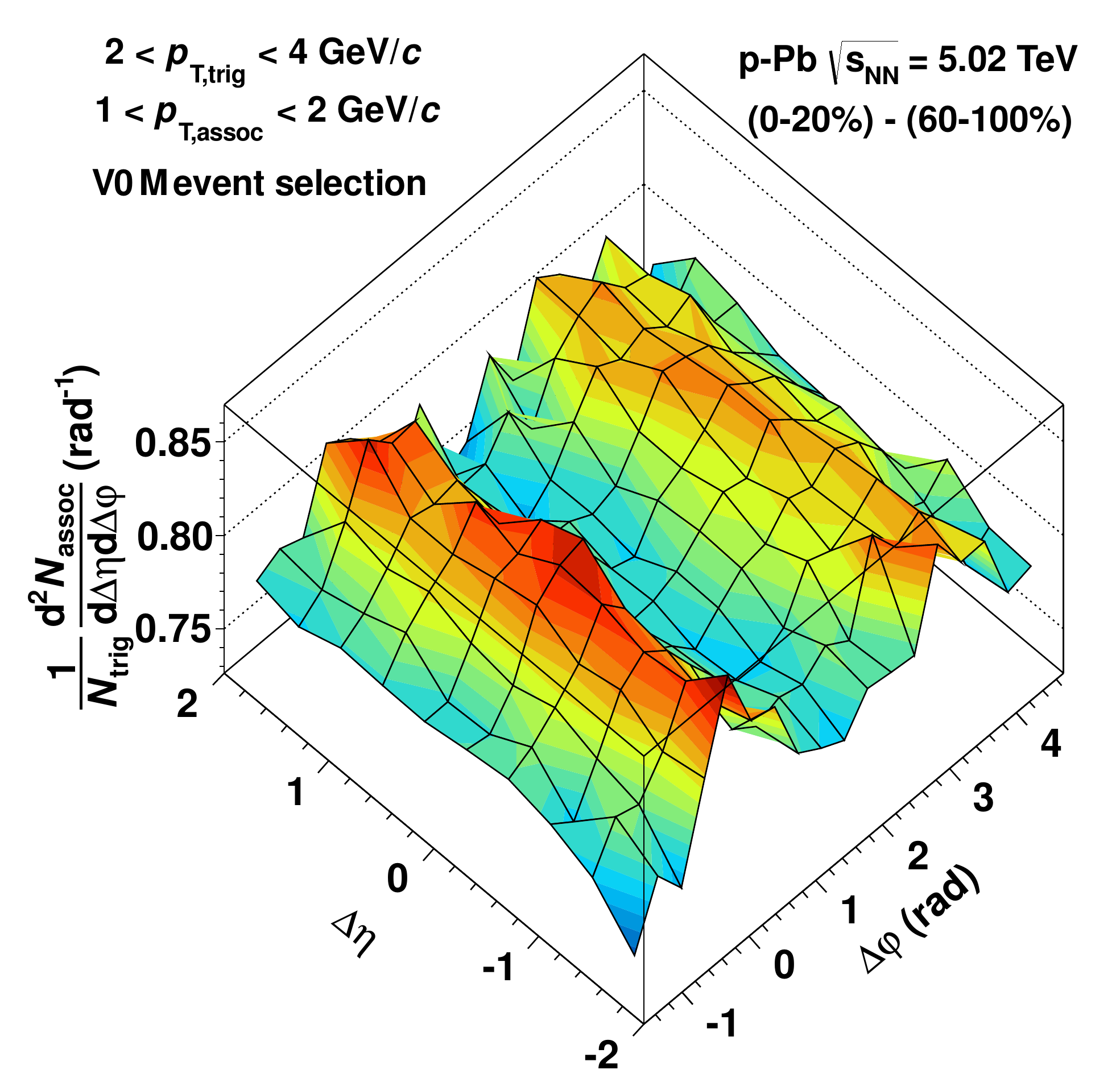}
\hspace{0.4cm}
\includegraphics[width=5.2cm,clip]{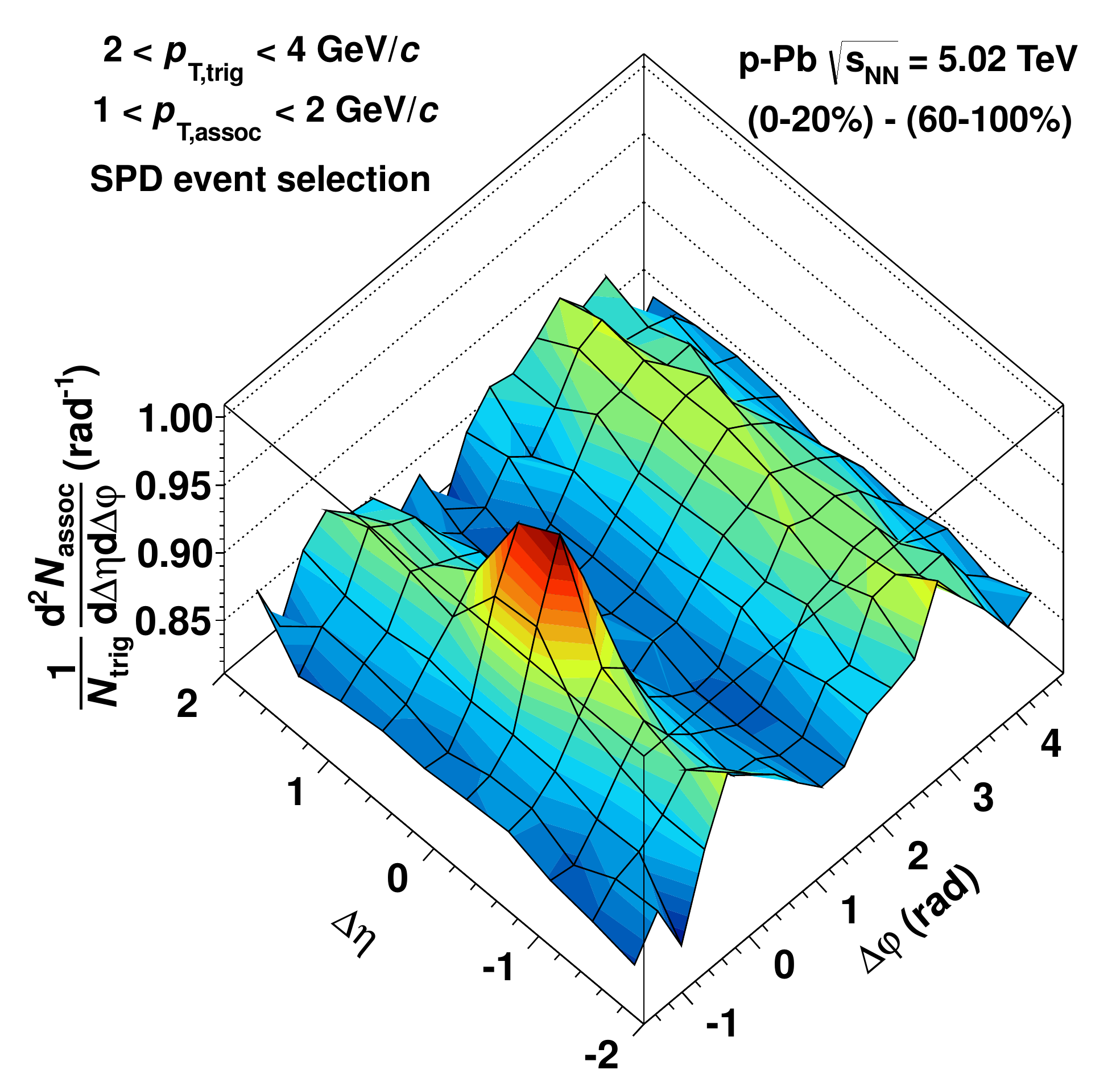}
\caption{Per-trigger particle associated yield in $\Dphi$ and $\Deta$ for pairs of charged particles with 
$2<\ptt<4$ GeV/$c$ and $1<\pta<2$ GeV/$c$ in \pPb\ collisions at $\snn=5.02$ TeV for the 0--20\% event class, 
selected in ZNA~(left panel), V0M~(middle panel) and SPD~(right panel), after subtraction of the yield obtained in the 
corresponding 60--100\% event class~\cite{Abelev:2012ola}.}
\label{fig:subalice}
\end{figure*}
The uncertainties related to tracking and trigger efficiency are regarded to be uncorrelated~(red boxes), 
while those related to the $\sqrt{s}$ dependence of \pp\ reference and $\avNcoll$ to be fully correlated~(gray bar),
and those related to the $y$ dependence of the \pp\ reference and the signal extraction to be partially correlated~(open boxes).
The data are compared to various calculations~(see the legend of \Fig{fig:jspirpa} for references).
Within the uncertainties, the models including shadowing or coherent energy loss are able to reproduce the data, 
while the prediction based on the Color Glas Condensate~(CGC) overestimates the observed suppression.

To be independent of the non-measured \pp\ reference the ratio~($\Rfb$) of J/$\psi$ produced symmetrically around $\yc=0$
in a forward over a backward rapidity interval is taken.
In this way, the uncertainties which are uncorrelated between backward and forward rapidity enter quadratically 
combined in the ratio, while for signal extraction the uncertainty can be directly calculated on the ratio of 
the number of signal events. For the forward-to-backward ratio, the main contribution to the uncertainty originates
from the tracking efficiency. Models including shadowing or coherent energy loss are qualitatively able 
to reproduce the $\Rfb$ measurement, either versus $\yc$ or versus $\pt$~(Figs.\ \ref{fig:jpsifby} 
and \ref{fig:jpsifbpt}; references for models can be found in the respective legends).

The measurements from~\cite{Aaij:2013zxa} indicate that cold nuclear matter effects are less pronounced 
for non-prompt than for prompt J/$\psi$, and are in agreement~\cite{ALICE:2013spa} for inclusive J/$\psi$ 
with those from \cite{Abelev:2013yxa}.

The results in \pPb\ indicate that the J/$\psi$ suppression observed in \PbPb\ collisions at 
$\snn=2.76$~TeV~\cite{Abelev:2012rv,Arnaldi:2012bg} can not be attributed to cold nuclear 
matter effects. However, firm conclusions will have to be drawn from model calculations.

\begin{figure}[htb!f]
\vspace{-4.8cm}
\centering
\includegraphics[height=11.6cm,clip,viewport=0 0 164 500]{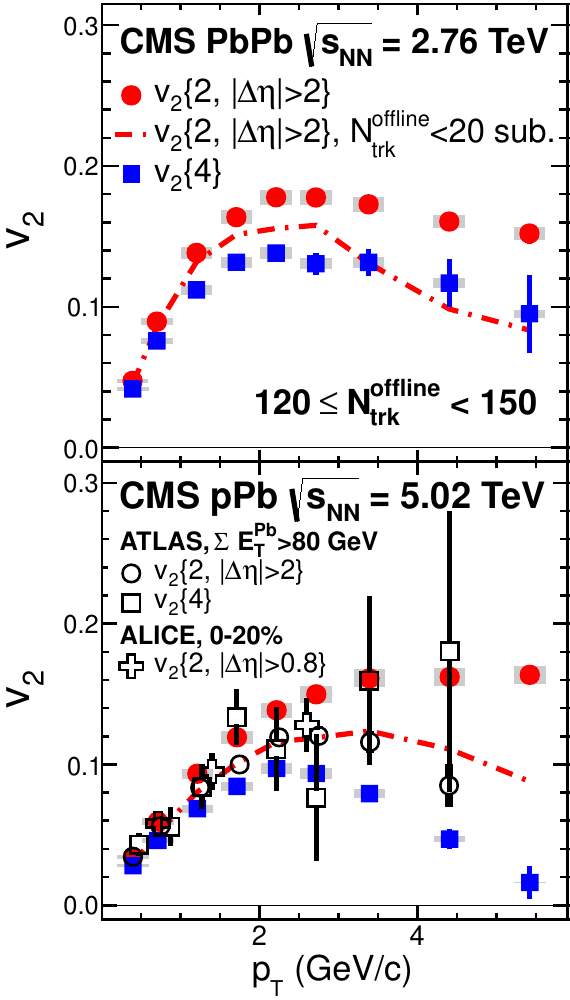}
\hspace{0.05cm}
\includegraphics[height=11.6cm,clip,viewport=0 0 165 500]{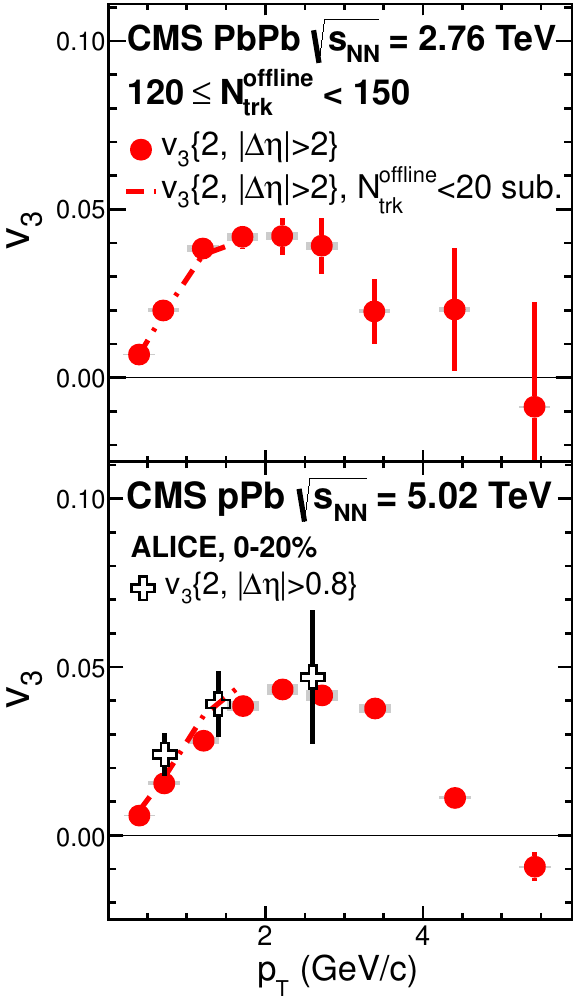}
\caption{
The $\pt$ differential $v_2$~(left panels) and $v_3$~(right panels) values in \PbPb\ collisions at $\snn=2.76$~TeV~(top panels) 
and \pPb\ collisions at $\snn=5.02$~TeV~(bottom panels) for one multiplicity interval. 
The results from CMS~\cite{Chatrchyan:2013nka} are $v_2\{2, |\Deta| > 2\}$ 
and $v_3\{2, |\Deta| > 2\}$~(filled circles) before and after subtracting the $\Ntrko<20$~(70--100\%) data~(dashed curves),
as well as $v_{2}\{4\}$~(filled squares) for the $120<\Ntrko<150$~(0--2\%) event class obtained with $|\etal|<2.4$ 
and $0.3<\ptref<3$~GeV/$c$.
The results from ALICE~\cite{Abelev:2012ola} are $v_2\{2, |\Deta| > 0.8\}$ and $v_3\{2, |\Deta| > 0.8\}$ (open cross)
obtained in $|\etal|<2$ for the 0--20\% subtracted by 60--100\% event class selected using forward-and-backward multiplicity~(V0M).
The results from \mbox{ATLAS}~\cite{Aad:2012gla,Aad:2013fja} are $v_2\{2, |\Deta| > 2\}$~(open circle) and $v_{2}\{4\}$~(open squares)
for the 0--2\% subtracted by the 50--100\% event class selected using the transverse energy on the Pb 
side~($\Sigma E^{\rm Pb}_{\rm T}$) obtained in $|\etal|<2.5$ and $0.3<\ptref<5$~GeV/$c$. 
The measurements~\cite{CMS:2012qk,Abelev:2012ola,Aad:2012gla,Aad:2013fja} use $2/\mu$b, while \cite{Chatrchyan:2013nka} uses $31/$nb.}
\label{fig:cmsv2v3}
\end{figure}

\section{Two- and four-particle correlations}
\label{correlations}
The study of angular correlations~(in $\phi$ and $\eta$) of two or more particles provides important information for the 
characterization of the underlying mechanism of particle production in collisions of hadrons and nuclei at high energy. 
For example, it is well known that in minimum-bias \pp\ collisions the correlation at ($\Dphi \approx 0$, $\Deta \approx 0$), the 
``near-side'' peak, and at $\Dphi \approx \pi$, the ``away-side'' structure, originates from particle production correlated to jets.
In \Aa\ collisions additional long-range structures along the $\Deta$ axis emerge on the near- and away-side, 
whose shape in $\Dphi$, typically quantified by Fourier coefficients $v_{n}$, can be related to the collision geometry 
and density fluctuations of the colliding nuclei in hydrodynamic models~(see~\cite{Heinz:2013th} for a review).
In \pp\ collisions at a centre-of-mass energy $\sqrt{s}=7$~TeV a similar long-range~(\mbox{$2<|\Deta|<4$}) structure, so called ridge,
emerges on the near-side in events with significantly higher-than-average particle multiplicity~\cite{Khachatryan:2010gv}.
Its origin has been attributed either to initial-state effects~(such as gluon saturation and colour connections forming along 
the longitudinal direction) or to final-state effects~(such as parton-induced interactions, and collective effects arising 
in a high-density system~(see \cite{Li:2012hc} for a review).

\begin{figure}[t!f]
\centering
\includegraphics[width=6.5cm,clip]{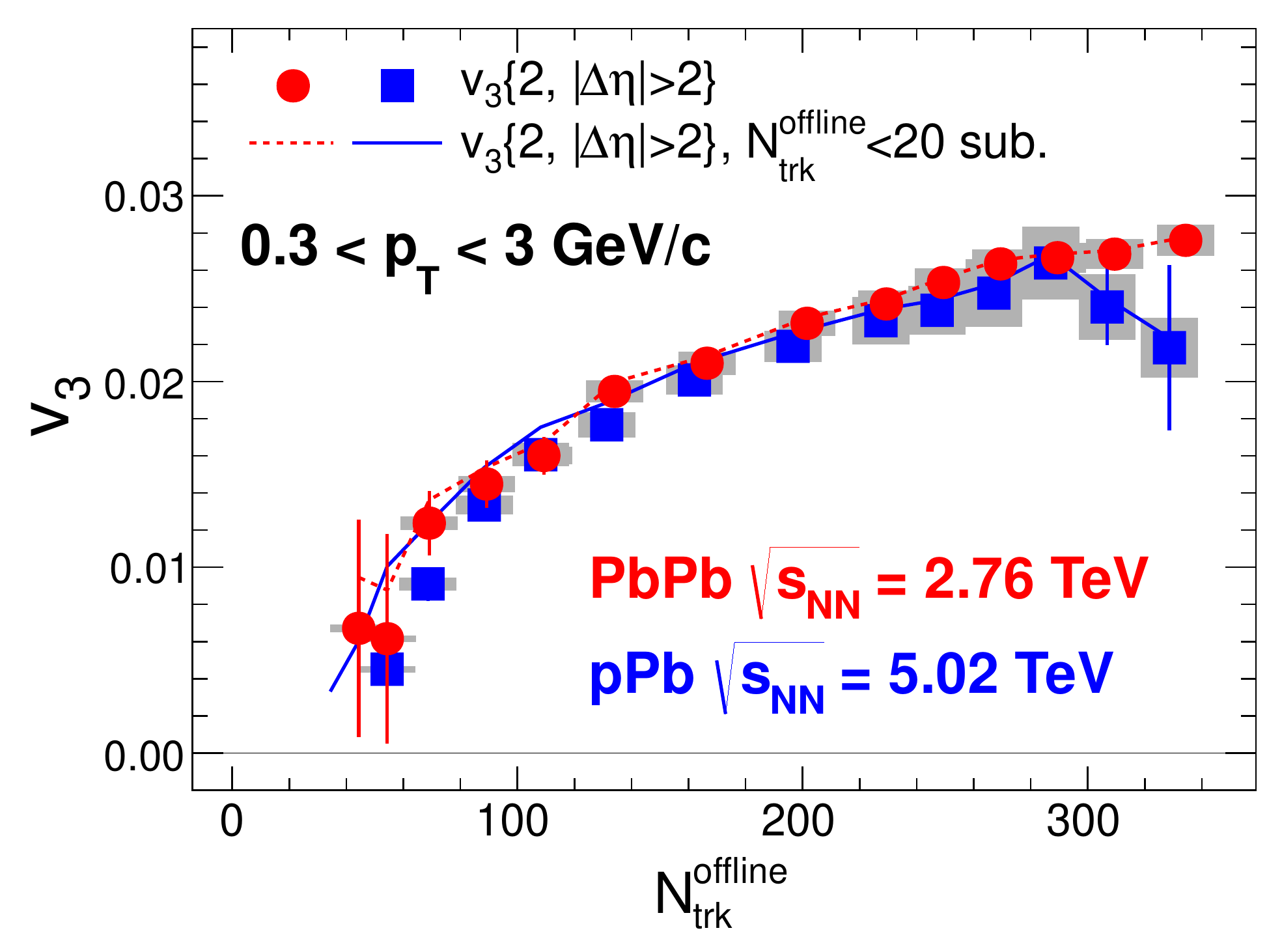}
\caption{
Integrated $v_3\{2, |\Delta\eta| > 2\}$ values as a function of $\Ntrko$ in \PbPb\ collisions at $\snn=2.76$~TeV and
in \pPb\ collisions at $\snn=5.02$~TeV. Details are given in the caption of \Fig{fig:cmsv2v3}.}
\label{fig:v3int}
\end{figure}

\begin{figure}[t!f]
\centering
\includegraphics[width=3.75cm,height=3.00cm,clip]{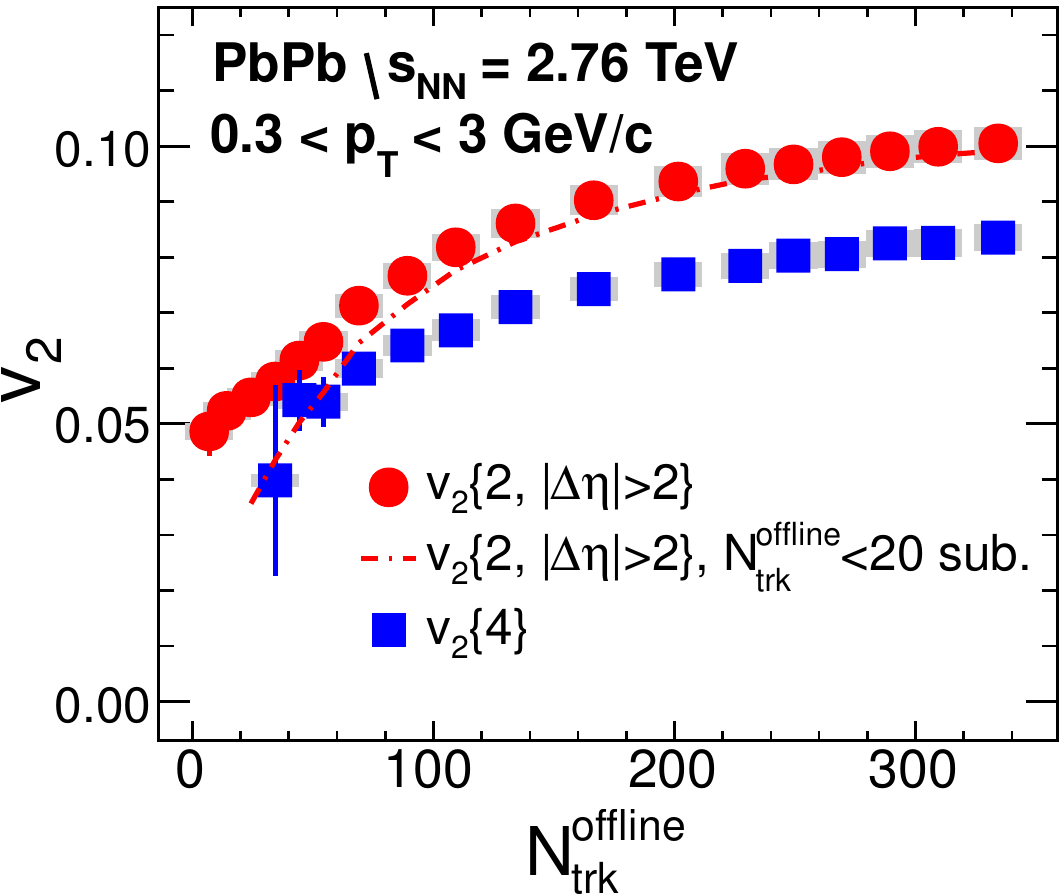}
\includegraphics[width=3.85cm,height=3.07cm,clip]{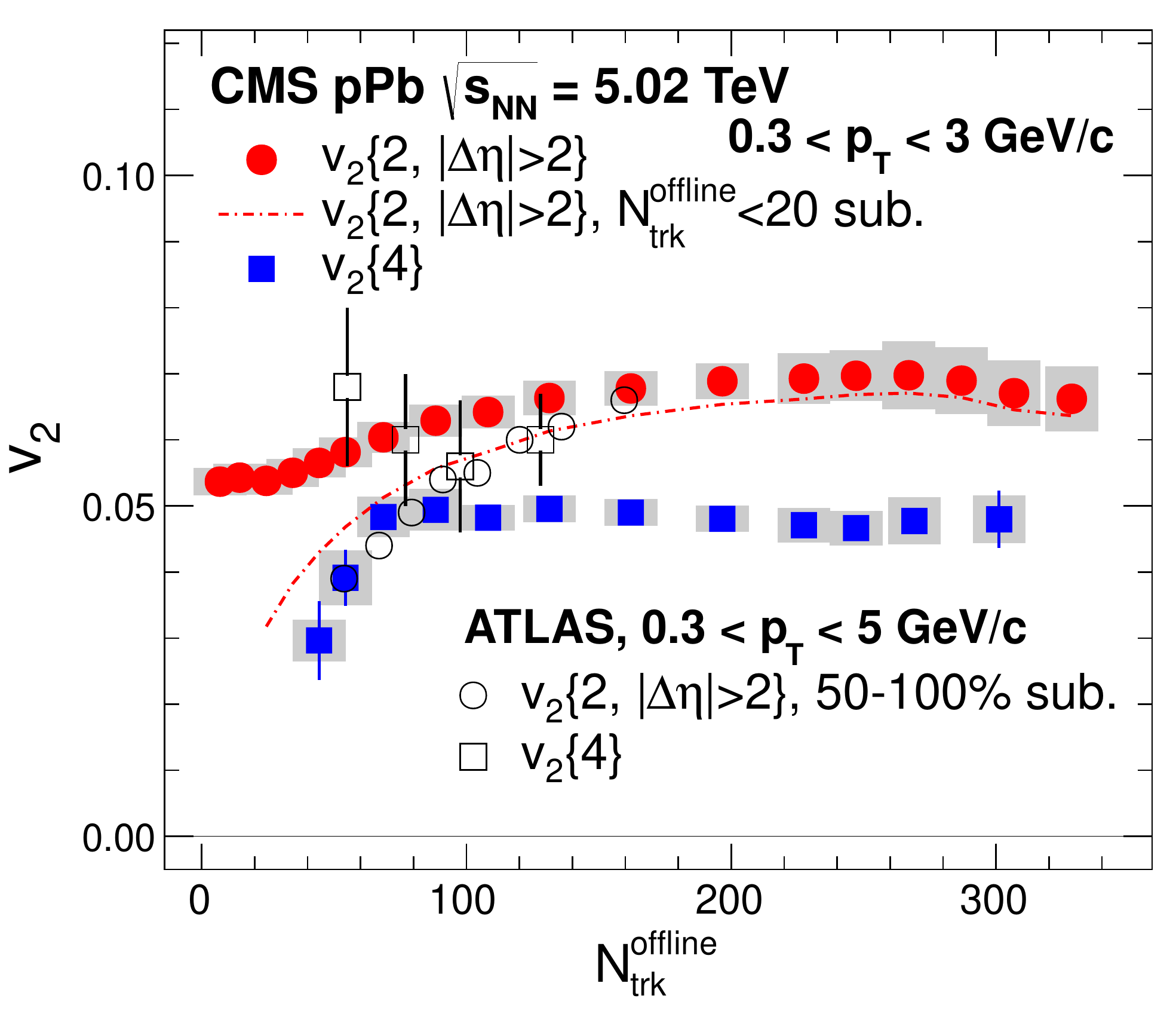}
\caption{
Integrated $v_2\{2, |\Delta\eta| > 2\}$, $v_2\{2, {\rm sub}\}$ and $v_{2}\{4\}$ values as a function of 
$\Ntrko$ in \PbPb\ collisions at $\snn=2.76$~TeV~(left panel) and in \pPb\ collisions at $\snn=5.02$~TeV~(right panel). 
Details are given in the caption of \Fig{fig:cmsv2v3}.} 
\label{fig:v2int}
\end{figure}

A qualitatively similar ridge, but with stronger correlation strength than in \pp, also appears on the near-side in high-multiplicity 
\pPb\ collisions at $\snn=5.02$~TeV~\cite{CMS:2012qk}. Subsequent measurements~\cite{Abelev:2012ola,Aad:2012gla}, which removed 
jet-induced correlations by subtracting the correlations extracted from low-multiplicity events, reveal that the near-side ridge 
is accompanied by essentially the same ridge on the away-side~(\Fig{fig:subalice}). 
Due to a bias on the jet production or fragmentation the subtraction of the jet peak is less complete, when the event selection is 
performed at midrapidity~(SPD) than at forward-and-backward rapidity~(V0M) or at beam rapidity~(ZNA).
The $\pt$ dependence of the extracted $v_2$ and $v_3$ coefficients from two-particle correlations is found to be similar to that 
measured in \PbPb\ collisions~(\Fig{fig:cmsv2v3}). 
This is in particular the case for $v_3$, where the $\pt$-integrated $v_3$ turns out to be the same in \pPb\ collisions and 
in \PbPb\ collisions at the same multiplicity~(\Fig{fig:v3int}). 
Differences between the two systems become apparent for $v_2\{4\}$, which is obtained by extracting the genuine four-particle 
correlations using cumulants~\cite{Bilandzic:2010jr}.
The integrated $v_2\{4\}$, as well as $v_2\{2\}$, are smaller (by up to about 35\%) than in \PbPb\ collisions at the same 
multiplicity~(\Fig{fig:v2int}). 

It is interesting to note that $v_2\{4\}$ and $v_3\{2\}$ set in at about the same multiplicity ($\Ntrko\approx50$ for $|\eta|<2.4$), 
which only is slightly larger than the average value for minimum bias \pPb\ collisions.
The interpretation of the correlation data focuses on two different approaches$\colon$ either quantum interference between 
rapidity-separated gluons enhanced by gluon saturation in the CGC model~\cite{Dusling:2012wy,Dusling:2013oia}, 
or collective dynamics induced by strong final-state interactions~\cite{Bozek:2012gr,Bozek:2013uha,Qin:2013bha},
as commonly applied in hydrodynamical models of \Aa\ collision data.
So far, the interpretation of the $v_3$ data is only achieved with the hydrodynamical approaches, and it is argued
that the observed effects result from rare fluctuations of the nucleon density~\cite{Coleman-Smith:2013rla}.
The similarity of the $v_2\{4\}$ and $v_3\{2\}$ data at fixed multiplicity in the \pPb\ and \PbPb\ systems can be explained
as the collective response to the fluctuations of clusters, when the geometrical contribution from the reaction plane in 
$v_2\{4\}$ is factorized out~\cite{Basar:2013hea}.
\ifarx
However, the application of hydrodynamics to a small system such as \pPb\ is complicated due to a significant
model dependence on the initial state geometry and its fluctuations and due to viscous corrections, which may 
be too large for hydrodynamics to be reliable~\cite{Bzdak:2013zma}.
\fi

\section{Identified particles}
\label{pidspectra}
Further experimental information expected to clarify whether final state effects play a role in high-multiplicity
\pPb\ collisions is provided by the measurement of identified particles.
So far, two measurements\co{ of identified particle $\pt$ spectra} are available$:$
\begin{itemize}
\item the $\pt$ spectra of \allpi, \allk\ and \allp\ measured by CMS via the energy loss in the silicon tracker
using $1/\mu$b for $|\yl|<1$ in the $\pt$ ranges of 0.1-1.2, 0.2-1.05 and 0.4-1.7~GeV/$c$, respectively, 
as a function of corrected track multiplicity~($\Ntracks$) in $|\etal|<2.4$~\cite{Chatrchyan:2013eya};
\item the $\pt$ spectra of \allpi, \allk, \kzero, \allp\ and \alll\ measured by ALICE via the energy loss in the
barrel tracking systems and via the time-of-flight information using $15/\mu$b for $0<\yc<0.5$
in the \pt\ ranges of 0.1-3, 0.2-2.5, 0-8, 0.3-4 and 0.6-8~GeV/$c$, respectively, 
as a function of midrapidity $\dNdeta$ selected in intervals of forward multiplicity~(V0A)~\cite{Abelev:2013haa}.
\end{itemize}

\begin{figure}[tbh!f]
\centering
\includegraphics[width=7.35cm,clip,viewport=0 0 360 345]{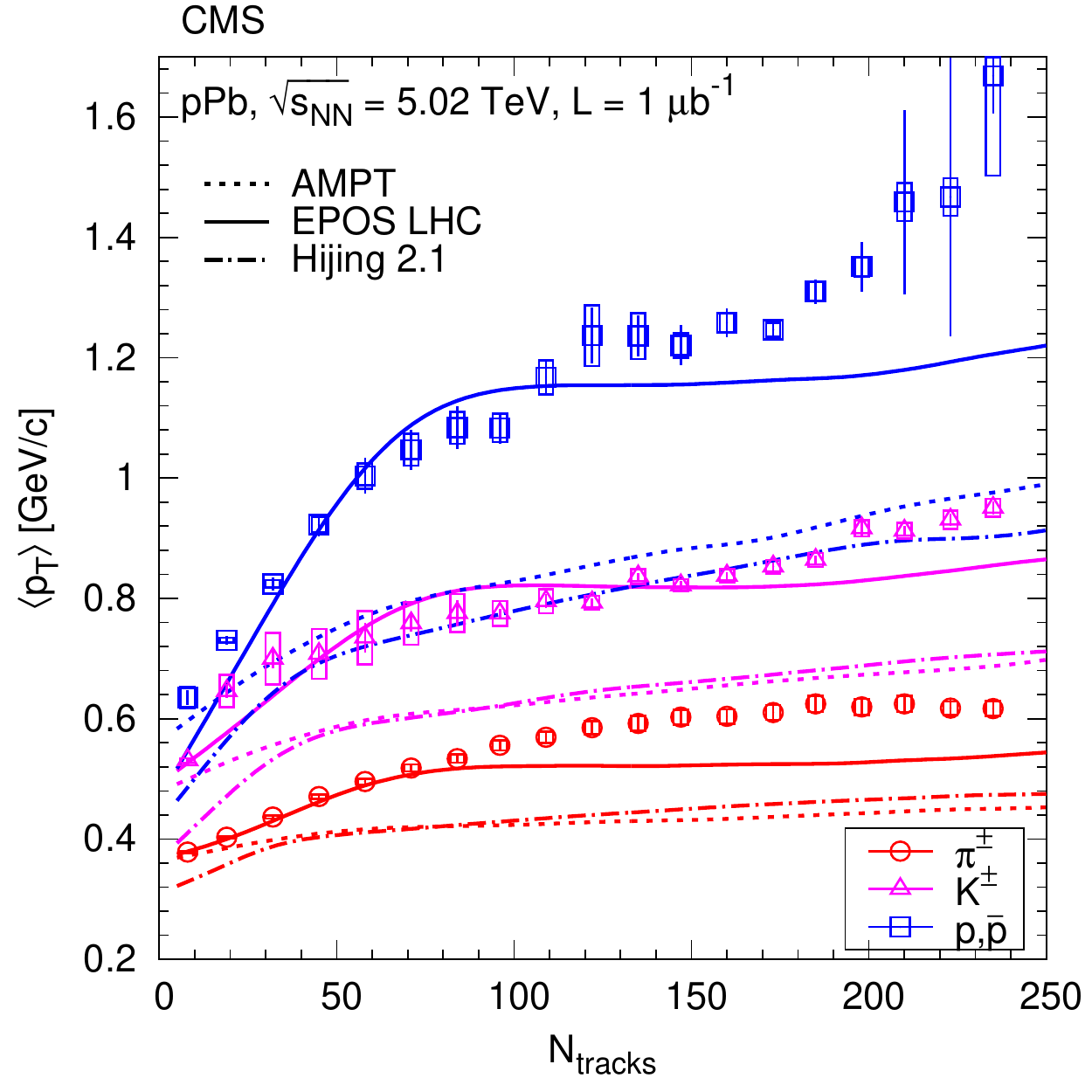}
\caption{
Average transverse momentum of \allpi, \allk\ and \allp\ in the range $|\yl|<1$ as a function of the corrected track
multiplicity for $|\etal|<2.4$ in \pPb\ collisions at $\snn=5.02$~TeV compared to model calculations~\cite{Chatrchyan:2013eya}.}
\label{fig:cmsmeanptpid}
\end{figure}
\begin{figure}[tbh!f]
\centering
\sidecaption
\includegraphics[width=4.4cm,height=3.5cm,clip,viewport=0 0 515 345]{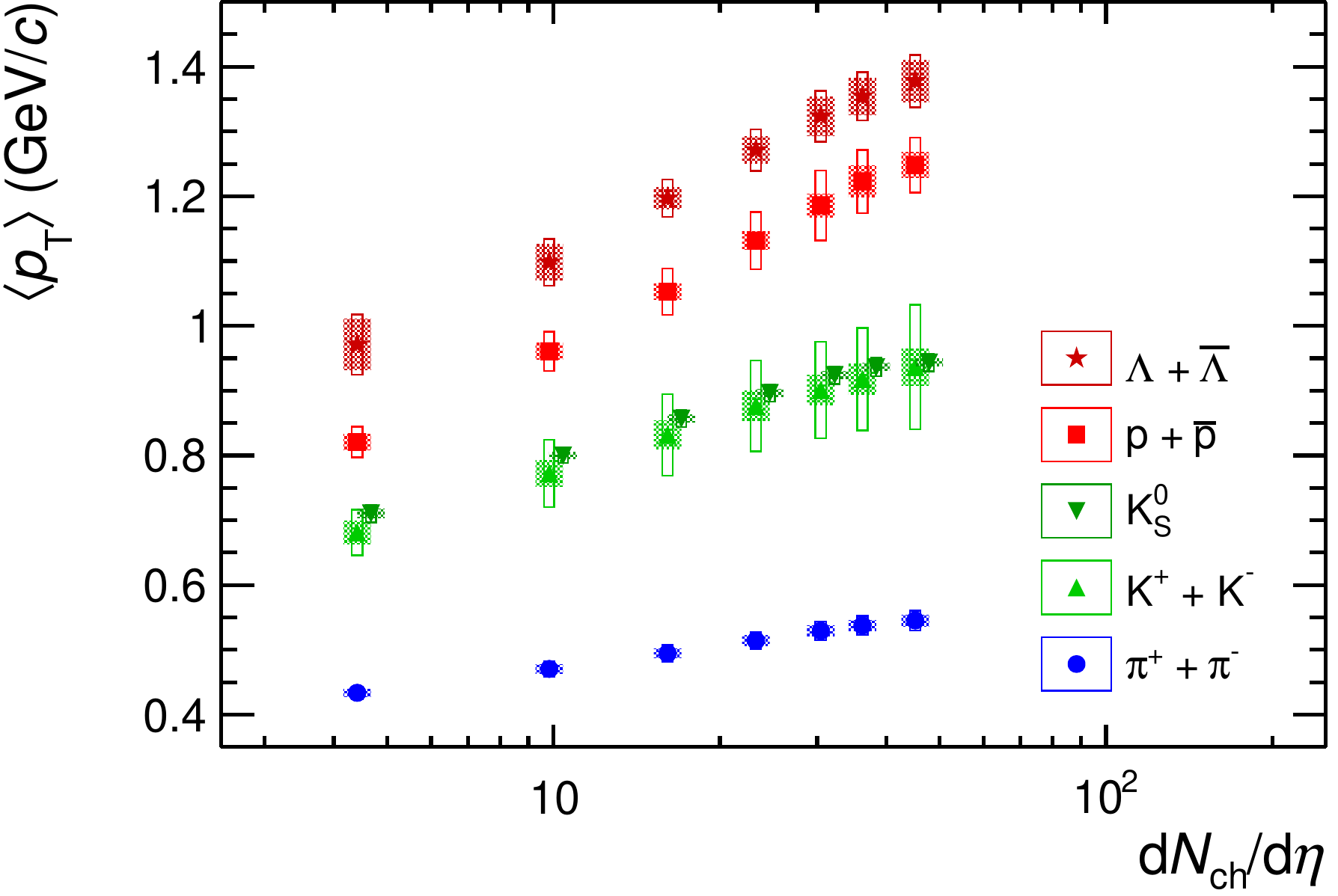}
\caption{
Average transverse momentum of \allpi, \allk, \kzero, \allp\ and \alll\ as a function of $\dNdetal$ 
in \pPb\ collisions at $\snn=5.02$ TeV~\cite{Abelev:2013haa}.
The $\dNdetal$ values of $\kzero$ are shifted for clarity.}
\label{fig:alicemeanptpid}
\end{figure}
\ifarx
\begin{figure}[tbh!f]
\centering
\includegraphics[width=7.35cm,clip,viewport=0 0 550 740]{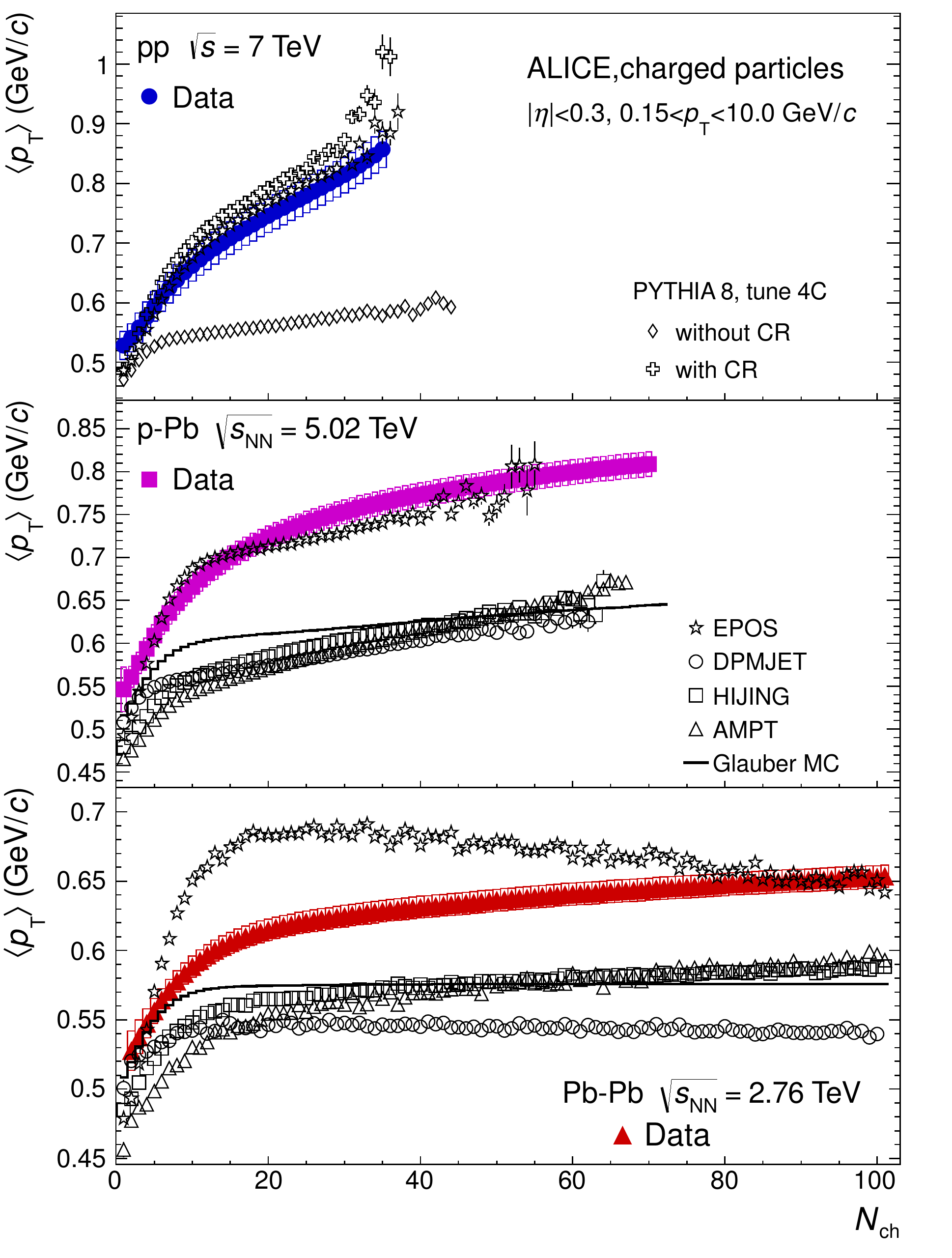}
\caption{Average transverse momentum as a function of $\Nch$ measured in \pp~(top panel), \pPb~(middle panel)
and \PbPb~(bottom panel) collisions compared to model calculations~\cite{Abelev:2013bla}.}
\label{fig:alicemeanpt}
\vspace{-1cm}
\end{figure}
\fi

To obtain the integrated yield and average $\pt$, the spectra are fitted for the extrapolation to zero and 
high $\pt$ in the unmeasured $\pt$ region.
In the case of CMS, a Tsallis-Pareto distribution is used with the unmeasured fraction of yield of about 
15–30\% for \allpi, 40–50\% for \allk, and 20–35\% for \allp.
In the case of ALICE, a blast-wave function is used with the unmeasured fraction of yield of about 8-9\% 
for \allpi, 10-12\% for \allk, 7-13\% for \allp\ and 17-30\% for \alll.
The measured $\pt$ spectra become harder for increasing multiplicity, with the change being most pronounced 
for \allp\ and \alll~(see Fig.~6 in \cite{Chatrchyan:2013eya} and Fig.~1 in \cite{Abelev:2013haa}).
This evolution is strongly reflected in the extracted average $\pt$, which is found to increase with particle mass and 
the charged multiplicity of the event~(\Fig{fig:cmsmeanptpid} and \Fig{fig:alicemeanptpid}).\ifarx~\footnote{The comparison 
of the average $\pt$ between the CMS and ALICE results is not straight forward because of the different selection bias
and $\eta$ ranges, as well as the different way the results are presented and extrapolated. If one nevertheless compares 
the data by taking $\Ntracks=4.8\,\dNdetal$, it turns out that \allp\ agrees within uncertainties except for first 
and last common data points, while the \allpi~(\allk) are systematically different between the two measurements.}\fi
This effect, called ``radial flow'' in a hydrodynamic scenario~\cite{Shuryak:2013ke}, is well known in \Aa\ 
collisions~(e.g.\ see \cite{Abelev:2013vea} for \PbPb\ at $\snn=2.76$~TeV).
Comparisons with calculations of Monte Carlo event generators reveal that EPOS LHC~\cite{Pierog:2013ria}, 
which (unlike HIJING or AMPT\co{references in \cite{Chatrchyan:2013eya}}) includes an hydrodynamic evolution of 
the created system, is able to reproduce the trend of the data~(\Fig{fig:cmsmeanptpid}). 

\ifarx
However, a qualitatively similar mass and multiplicity dependence of the identified-particle average $\pt$ has been 
also found in \pp\ collisions at the LHC~\cite{Chatrchyan:2012qb}. It has been demonstrated that color string formation 
between final partons from independent hard scatterings, called ``color reconnection''~(CR)~\cite{Sjostrand:2013cya}, 
can mimic the ``flow-like'' trends seen in the \pp\ data~\cite{Ortiz:2013yxa}. 
Unlike hydrodynamics, the CR mechanism acts on a microscopic level, and therefore does not require the formation
of a (partially) thermalized medium in a small system.

Similar conclusions can be drawn from a measurement of the average $\pt$ for charged particles 
with $0.15<\pt<10$~GeV/$c$ in $|\etac|<0.3$ as a function of the number of charged particles with $\pt>0$ 
in $|\etac|<0.3$~($\Nch$), compared between \pp, \pPb, and \PbPb\ collisions at the same $\Nch$~(\Fig{fig:alicemeanpt}).
The CR mechanism describes the increase of the average $\pt$ with $\Nch$ in \pp\ collisions. 
Models that are only based on a superposition of independent nucleon--nucleon collisions 
fail to describe the average $\pt$ in \pPb~(and \PbPb) collisions.
It is argued~\cite{McLerran:2013oju,Rezaeian:2013woa} that the dependence of average $\pt$ for identified particles in \pp\ and \pPb\ 
collisions with multiplicity can be generally described using geometric scaling and the dependence on the transverse 
interaction area with multiplicity as computed in the CGC framework. However, this scaling is found to hold less well 
in the case of unidentified particles~\cite{Abelev:2013bla}.
\fi

\ifarx
\begin{figure}[tbh!f]
\centering
\includegraphics[width=6.5cm,clip,viewport=0 0 550 350]{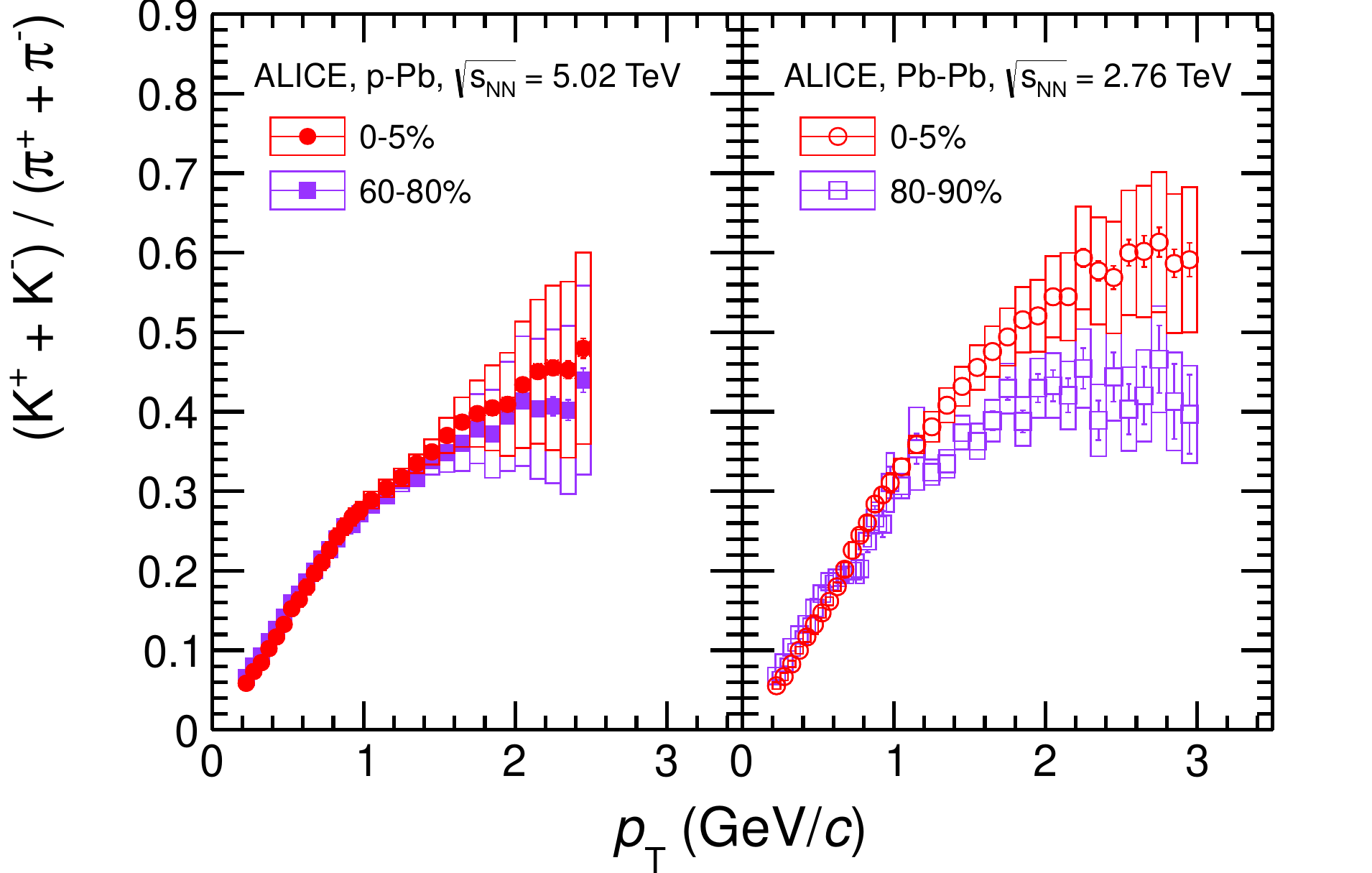}
\vfill
\vspace{-0.42cm}
\includegraphics[width=6.5cm,clip,viewport=0 0 550 350]{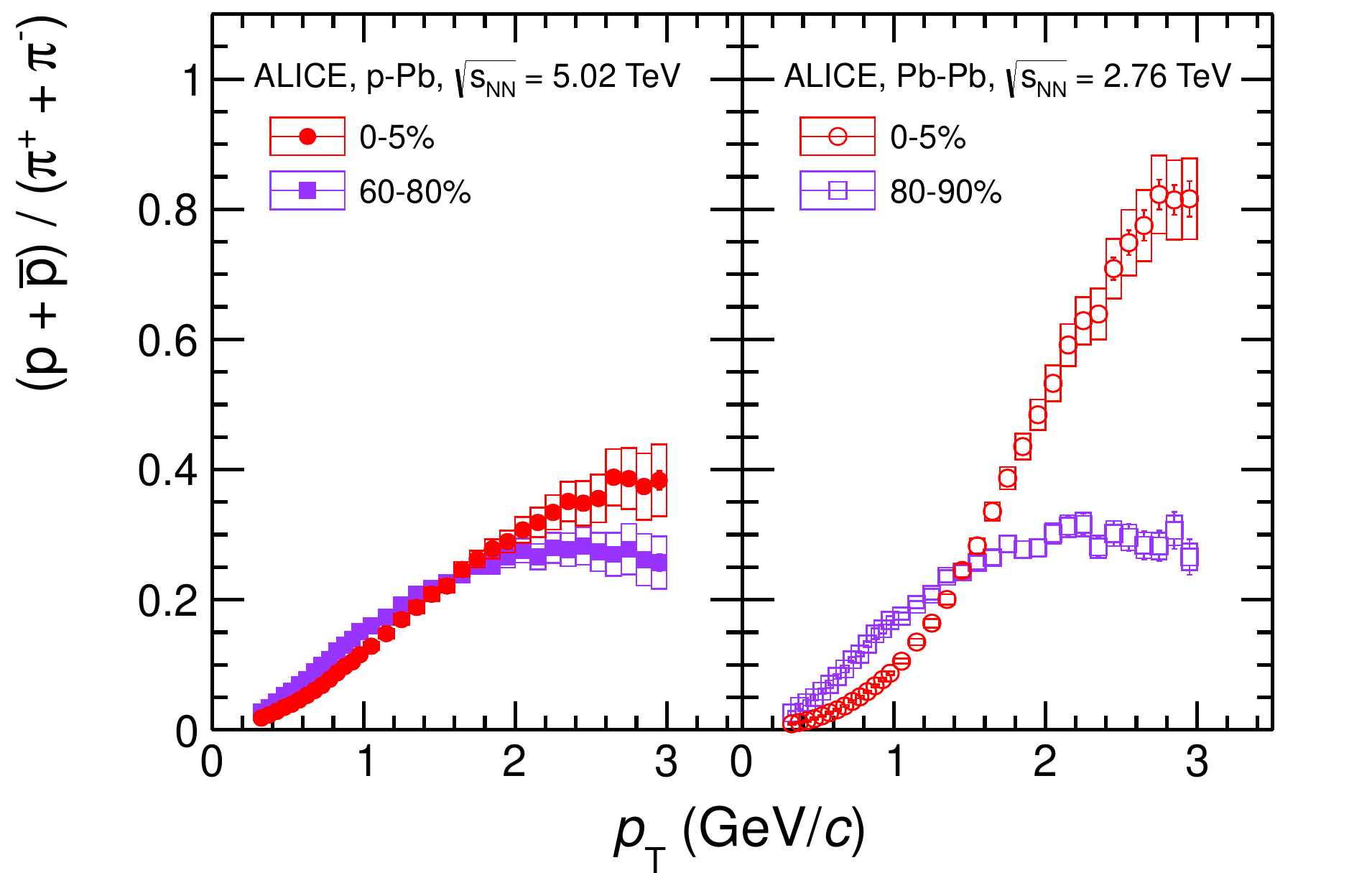}
\vfill
\vspace{-0.42cm}
\includegraphics[width=6.5cm,clip,viewport=0 0 550 350]{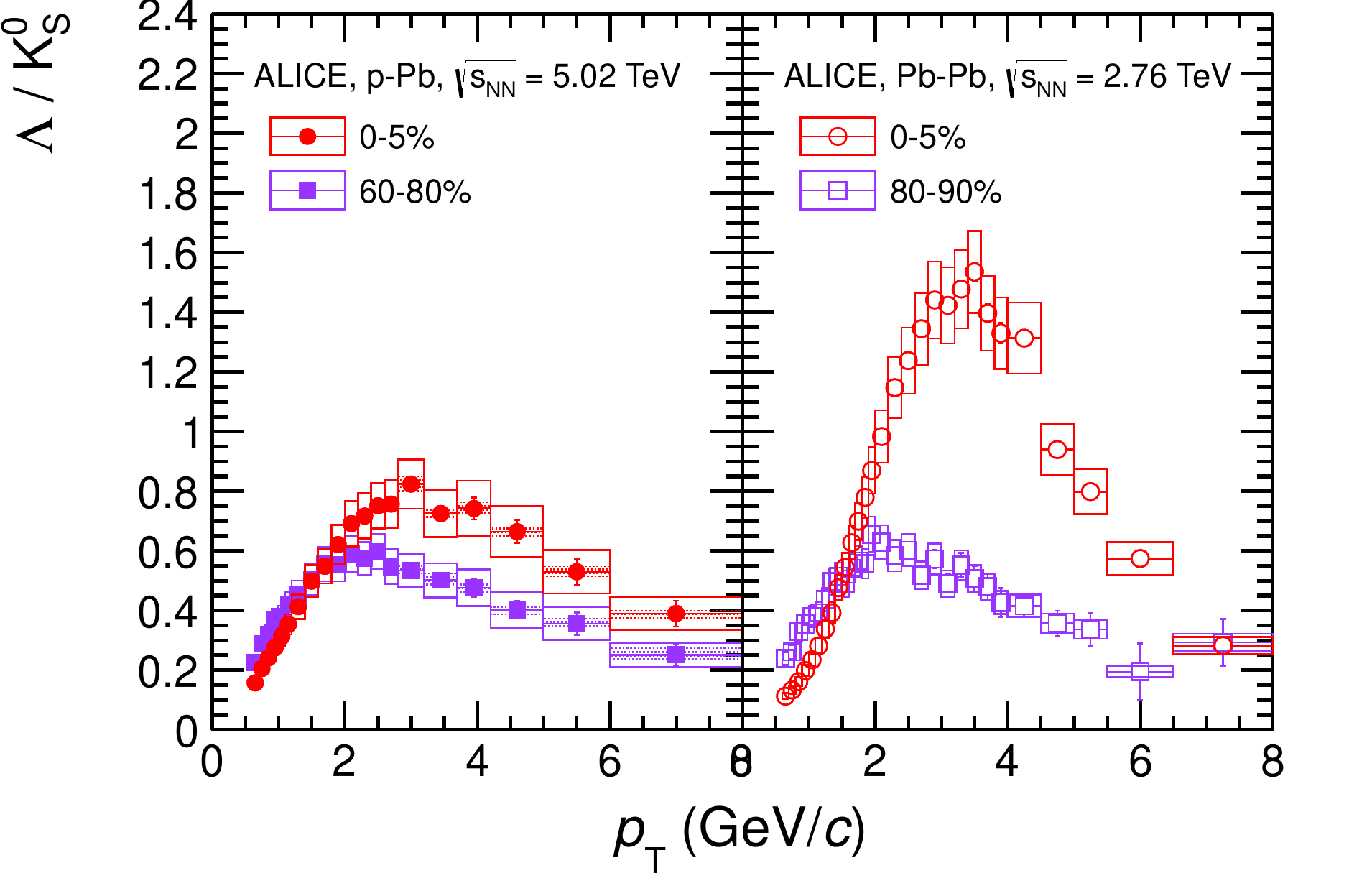}
\caption{
Ratios $\rm K/\pi$, $\rm p/\pi$, and $\rm \Lambda/\kzero$ as a function of \pt\ in \pPb\ collisions at 
$\snn=5.02$~TeV~(left panels) compared to those in \PbPb\ collisions at $\snn=2.76$~TeV in two event
classes~\cite{Abelev:2013haa}. 
The \pPb\ event classes are determined using V0A, while the \PbPb\ classes using V0M.  
}
\label{fig:ratios}
\end{figure}
\fi

\ifarx
\begin{figure}
\centering
\sidecaption
\includegraphics[width=7.5cm,clip]{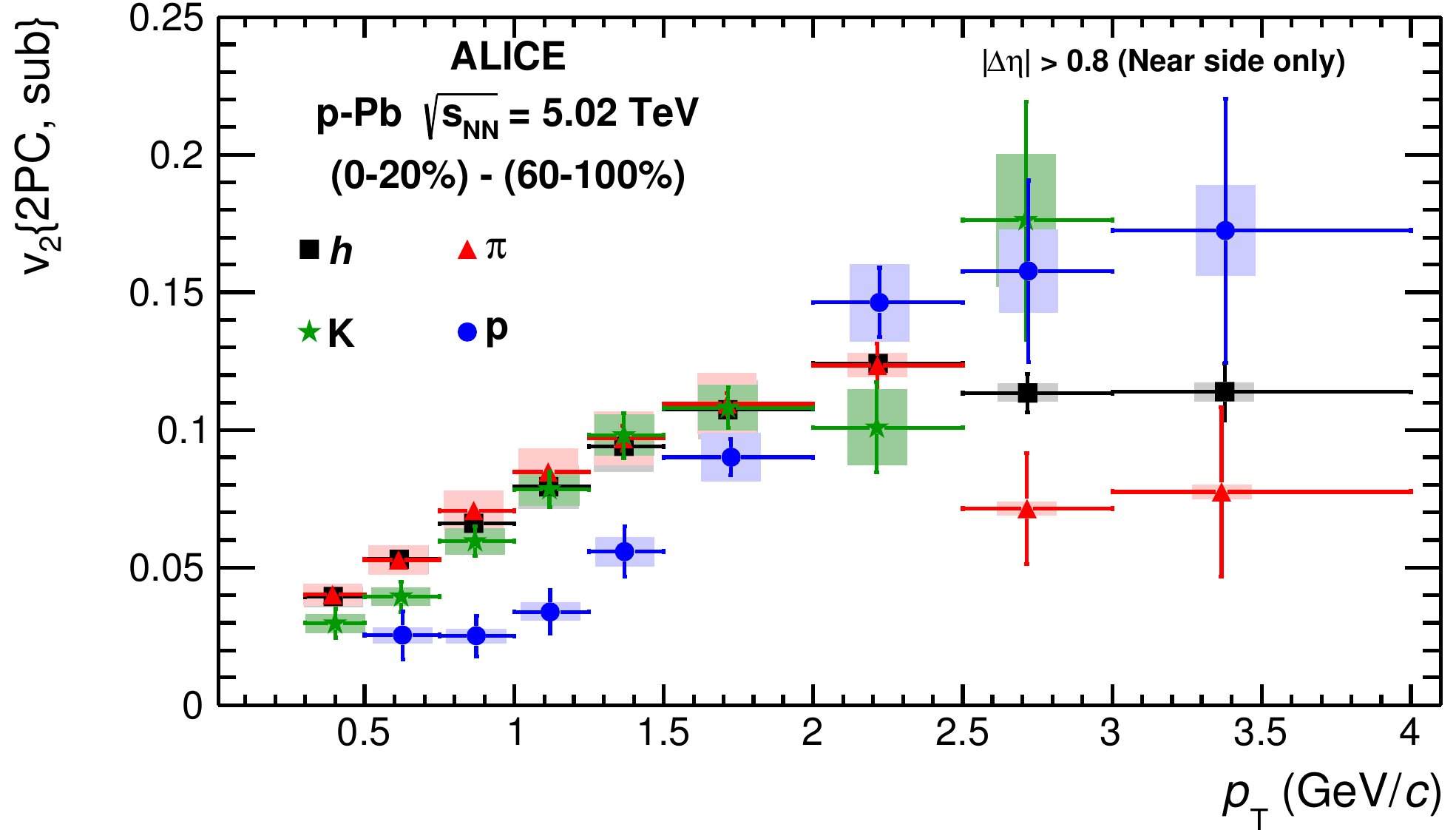}
\caption{
The $v_2$\co{$v_2\{{\rm 2PC,sub}\}$} values extracted from two-particle correlations in \pPb\ collisions at $\snn=5.02$~TeV
for hadrons (black squares), pions (red triangles), kaons (green stars) and protons (blue circles) as a function of 
$\pt$ in the 0--20\% after subtraction of the 60--100\% event class~(selected with V0A)~\cite{Abelev:2013wsa}. 
The data are plotted at the average-$\pt$ for each $\pt$ interval.
}
\label{fig:v2pid}
\end{figure}
\fi

\ifarx
Additional information is obtained by comparing the ${\rm K}/\pi$, ${\rm p}/\pi$ and ${\rm \Lambda}/\pi$
ratios as a function of $\pt$  between \pPb\ and \PbPb\ collisions for two event classes~(\Fig{fig:ratios}).
The ${\rm p}/\pi$ and ${\rm \Lambda}/\pi$ ratios exhibit a significant enhancement at intermediate $\pt\approx3$~GeV/$c$,
qualitatively similar to what is observed in \PbPb\ collisions~\cite{Abelev:2013vea, Abelev:2013xaa}.  
The magnitude of the observed effects, however, differs between the \pPb\ and \PbPb\ systems, with peripheral
\PbPb\ collisions\co{($\dNdetal\approx13$)} roughly resembling the highest multiplicity class in \pPb\co{($\dNdeta\approx45$)}.
The observations in \PbPb\ are typically attributed to collective flow or quark recombination~\cite{Muller:2012zq},
while similar observations in \pp\ could originate from the CR mechanism~\cite{Ortiz:2013yxa}.

In \PbPb\ collisions, the $v_2$ for identified particle species exhibits a characteristic particle-mass dependent
splitting, with the $v_2$ of lighter identified particles found to be larger than that of heavier particles 
at the same $\pt$~\cite{Noferini:2012ps}. The splitting can be understood in the presence of a collective expansion, 
as for example predicted by hydrodynamic model calculations~\cite{Shen:2011eg}.
In \pPb\ collisions a similar mass dependent splitting of the $v_2$ coefficients is observed, with the 
$v_2$ of \allp\ being significantly lower than that of \allpi and \allk~(\Fig{fig:v2pid}).
The $v_2$ values are extracted from the per-trigger yield of identified associated particles\co{(\allpi, \allk\ or \allp)}
relative to charged trigger particles in symmetric intervals of $\pt$ in the 0--20\% event class~(selected with V0A) 
after subtracting the per-trigger yields from the 60--100\% event class using 50/$\mu$b.
As in the case of \PbPb, the splitting can be described by hydrodynamical model calculations~\cite{Bozek:2013ska,Werner:2013ipa}.
It should be noted that a microscopic mechanism as the fore-mentioned CR might create a similar effect, which however has
not yet been investigated.
\fi

\section{Summary}
\ifarx
\label{summary}The first results~\cite{ALICE:2012xs,ALICE:2012mj,Chatrchyan:2014hqa,Abelev:2013yxa,Aaij:2013zxa,CMS:2012qk,
Abelev:2012ola,Aad:2012gla,Aad:2013fja,Chatrchyan:2013nka,Chatrchyan:2013eya,Abelev:2013haa,Abelev:2013wsa,Abelev:2013bla}
from \pPb\ collisions at \snn = 5.02 TeV are discussed.
The strong hadron suppression and dijet momentum imbalance seen for (central) \PbPb\ collisions at $\snn=2.76$~TeV 
can not be attributed to modification of the initial state alone.
The data on J/$\psi$ production are reproduced by models including shadowing or coherent energy loss.
At similar multiplicity, the \pPb\ spectra and azimuthal correlation data, as well as the \pp\ spectra at $\sqrt{s}=7$~TeV, 
exhibit characteristic features that are qualitatively similar to those from \PbPb\ collision at $\snn=2.76$~TeV.
These features can all be understood assuming the presence of final state interactions, as maximally realized in hydrodynamical 
models, which were originally developed to explain the \PbPb\ data. 
A microscopic approach such as the color reconnection mechanism relevant for the \pp\ data may provide an alternative explanation 
without having to rely on hydrodynamics.
Some of observations are also described by quantum interference effects computed in the CGC framework.

\section*{Acknowledgements}
\else
\label{summary}The first results from \pPb\ collisions at \snn = 5.02 TeV are discussed.
\fi
I would like to thank the members of the ALICE, \mbox{ATLAS}, CMS and LHCb collaborations for the use of their data, 
and the CERN accelerator team for the excellent \pPb\ operations.

%
\bibliography{biblio}{}
%
%
%
%
\end{document}